\documentclass[prd,amsmath,amssymb,aps,superscriptaddress,twocolumn,nofootinbib
]{revtex4-2}

\usepackage{epsfig}
\usepackage{url}
\usepackage{hyperref}

\usepackage[normalem]{ulem} 

\usepackage{latexsym}
\usepackage{epsfig}
\usepackage{amsmath}
\usepackage{amssymb}
\usepackage{wasysym}
\usepackage{graphicx}
\usepackage{dcolumn}
\usepackage{verbatim}
\usepackage{enumerate,mdwlist}

\usepackage[titletoc]{appendix}

\usepackage{amsfonts}
\usepackage{fancyvrb} 
\usepackage{tikz} 
\usetikzlibrary{calc}
\usepackage[export]{adjustbox}

\usepackage[normalem]{ulem}

\usepackage[inline, final]{showlabels}
\usepackage{tensor}

\usepackage{listings}
\usepackage{xcolor}
\usepackage{fontawesome5}

\definecolor{comment_red}{rgb}{0.5, 0, 0}
\newcommand{\ee}{\text{e}} 
\newcommand{\ii}{\text{i}}
\newcommand{\di}{\text{d}}

\makeatletter
\g@addto@macro\bfseries{\boldmath}  
\makeatother

\renewcommand{\v}[1]{\boldsymbol{#1}}

\renewcommand{\d}[2]{\frac{\text{d} #1}{\text{d} #2}}

\newcommand{\ba}{\begin{eqnarray}}
\newcommand{\ea}{\end{eqnarray}}

\newcommand{\glow}{\texttt{GLoW}}

\definecolor{grey}{rgb}{0.4,0.4,0.4}
\definecolor{dullmagenta}{rgb}{0.4,0,0.4}
\definecolor{darkblue}{rgb}{0,0,0.4}
\definecolor{midblue}{rgb}{0,0,0.5}
\definecolor{midred}{rgb}{0.5,0,0}
\definecolor{orange}{rgb}{1,0.5,0}
\definecolor{lightbrown}{rgb}{0.75,0.5,0.25}
\definecolor{tan}{cmyk}{0.14,0.42,0.56,0}
\definecolor{djunglegreen}{cmyk}{0.99,0,0.52,0}
\definecolor{lightgreen}{rgb}{0,1,0}
\definecolor{olivegreen}{cmyk}{0.64,0,0.95,0.40}
\definecolor{midgreen}{rgb}{0.0,0.675,0.0}
\definecolor{darkgreen}{rgb}{0,0.5,0}
\definecolor{ceruleanblue}{rgb}{0.0, 0.2, 0.7}
\definecolor{burgundy}{rgb}{0.5, 0.0, 0.13}
\definecolor{hvred}{RGB}{186,12,47}

\hypersetup{
    colorlinks=true,
    linkcolor=ceruleanblue,
    filecolor=ceruleanblue,      
    urlcolor=midblue,
    citecolor=burgundy,
}

\usepackage[all]{xy} 
\usepackage{amsfonts}

\makeatletter
\def\l@subsubsection#1#2{}
\makeatother

\begin{document} 

\title{GLoW: novel methods for wave-optics phenomena in gravitational lensing}

\author{Hector Villarrubia-Rojo}
\email{hectorvi@ucm.es}
\affiliation{Departamento de Física Teórica, Universidad Complutense de Madrid,
28040 Madrid, Spain}
\affiliation{Max Planck Institute for Gravitational Physics (Albert Einstein Institute) \\
Am Mühlenberg 1, D-14476 Potsdam-Golm, Germany}

\author{Stefano Savastano}
\email{stefano.savastano@aei.mpg.de}
\affiliation{Max Planck Institute for Gravitational Physics (Albert Einstein Institute) \\
Am Mühlenberg 1, D-14476 Potsdam-Golm, Germany}

\author{Miguel Zumalac\'arregui}
\email{miguel.zumalacarregui@aei.mpg.de}
\affiliation{Max Planck Institute for Gravitational Physics (Albert Einstein Institute) \\
Am Mühlenberg 1, D-14476 Potsdam-Golm, Germany}

\author{Lyla Choi}
\email{lc3535@princeton.edu}
\affiliation{Department of Physics, Princeton University, Princeton, NJ 08544}

\author{Srashti Goyal}
\email{srashti.goyal@aei.mpg.de}
\affiliation{Max Planck Institute for Gravitational Physics (Albert Einstein Institute) \\
Am Mühlenberg 1, D-14476 Potsdam-Golm, Germany}


\author{Liang Dai}
\affiliation{University of California at Berkeley,
Berkeley, California 94720, USA}

\author{Giovanni Tambalo}
\email{gtambalo@phys.ethz.ch}
\affiliation{Institut f\"ur Theoretische Physik, ETH Z\"urich, 8093 Z\"urich, Switzerland}

\begin{abstract}
Wave-optics phenomena in gravitational lensing occur when the signal's wavelength is commensurate to the gravitational radius of the lens. Although potentially detectable in lensed gravitational waves, fast radio bursts and pulsars, accurate numerical predictions are challenging to compute.
Here we present novel methods for wave-optics lensing that allow the treatment of general lenses. In addition to a general algorithm, specialized methods optimize symmetric lenses (arbitrary number of images) and generic lenses in the single-image regime. We also develop approximations for simple lenses (point-like and singular isothermal sphere) that drastically outperform known solutions without compromising accuracy. 
These algorithms are implemented in \textit{Gravitational Lensing of Waves} (\glow): an accurate, flexible, and fast code.
\glow{} efficiently computes the frequency-dependent amplification factor for generic lens models and arbitrary impact parameters in $\mathcal{O}$(1 ms)  to $\mathcal{O}$(10 ms) depending on the lens configuration and complexity.
\glow{} is readily applicable to model lensing diffraction on gravitational-wave signals, offering new means to investigate the distribution of dark-matter and large-scale structure with signals from ground and space detectors.
\end{abstract}

\date{\today}




\maketitle
{
  \hypersetup{hidelinks}
  \tableofcontents
}

\section{Introduction}\label{sec:intro}

Like a magnifying glass, gravitational fields deflect and focus signals propagating through the universe, in some cases producing multiple images of the same source~\cite{Schneider1999-tk}. This broad set of phenomena, collectively denoted \textit{gravitational lensing}, is essential to interpret many astronomical observations correctly. In addition, observation of lensing effects has led to many applications in cosmology and astrophysics, which include finding extra-solar planetary systems~\cite{Tsapras_exoplanets_microlensing_review,Gaudi:2008zq}, directly imaging supermassive black holes~\cite{EventHorizonTelescope:2019dse,EventHorizonTelescope:2022wkp}, mapping the distribution of dark matter~\cite{Clowe:2006eq,Planck:2018lbu,Vegetti:2023mgp}, measuring the Universe's expansion~\cite{H0LiCOW:2019pvv} and testing gravity~\cite{Reyes:2010tr,Collett:2018gpf}.

In most applications, gravitational lensing can be understood as signals propagating over well-defined trajectories, denoted rays~\cite{Leung:2023lmq}. Rays define the image positions, their magnifications, relative arrival times, and apparent deformation of a source. 
This description, known as \textit{geometric optics} (GO), emerges through Fermat's principle, stating that light travels between two points along the path that requires the least time. 
Geometric optics is valid for signals with wavelengths much smaller than the difference in path length. This is satisfied in all electromagnetic lensing phenomena observed so far, due to the large hierarchy between typical astronomical objects and the radiation able to penetrate the Earth's atmosphere. 

The \textit{wave optics} regime describes the propagation of signals with arbitrarily low frequencies, where geometric optics fails. The primary effect observed in this regime is \textit{diffraction}, the wavefront distortion caused by obstacles during propagation. Diffraction is responsible for frequency-dependent effects on the signal. It is well-known in optics, where diffraction patterns have found many applications, e.g. to characterize materials (crystallography), but these uses rely on light-matter interactions. Gravitational diffraction of electromagnetic signals can be caused by extremely light objects. Although it has not yet been observed, electromagnetic lensing diffraction could be used to probe planet-scale objects and compact dark-matter objects~\cite{Ulmer:1994ij,Oguri:2019fix, Jow:2020rcy,Tamta:2024pow}. 


Gravitational waves (GWs) offer a promising observational window for wave-optic lensing phenomena~\cite{Takahashi:2003ix}. Phase coherence of GWs prevents the blurring of diffraction patterns expected in extended sources, and ab initio emission models can help discern frequency-dependent distortions of the signal.  Typical GW wavelengths are orders of magnitude larger than radio waves that can penetrate the atmosphere, making diffraction by typical lenses observable. GWs in ground-based detectors ($\sim 100$ Hz)  are sensitive to diffraction from objects with $1-10^3 M_\odot$ (solar masses)~\cite{Christian:2018vsi}, including optically thick stellar populations in galaxies producing strong lensing ~\cite{Diego:2019lcd,Mishra:2021xzz,Shan:2023ngi}.
Detectors at lower frequencies can probe much heavier structures: space detectors like LISA (mHz) can probe halos/subhalos with $10^5-10^8M_\odot$~\cite{Caliskan:2023zqm,Brando:2024inp} and pulsar-timing arrays (nHz) can probe wave-optics by galactic-scale objects $\sim 10^{12}M_\odot$~\cite{Jow:2024bwq}. 
The prospect of detection and potential applications requires the development of new tools to explore wave-optics lensing phenomena~\cite[Sec.~11.2]{LISACosmologyWorkingGroup:2022jok}.

Accurately computing wave-optics lensing predictions is numerically challenging.
General predictions require conditionally convergent integrals of rapidly oscillating functions over the lens plane. Previous studies have used direct integration~ \cite{Takahashi:2004mc,Dai:2018enj}, Levin's method~\cite{Moylan:2007fi, Guo:2020eqw}, sampling the Fermat potential over isochrone lines~\cite{Ulmer:1994ij,Mishra:2021xzz,cheung2024probingminihalolensesdiffracted}, by discretizing the lens plane~\cite{Diego:2019lcd, Cheung:2020okf, Yeung:2021roe}, by direct fast-fourier transform convolution~\cite{Grillo:2018qjt}, or using analytic continuation (Picard-Lefschetz) theory~\cite{Feldbrugge:2019fjs,Jow:2022pux}. 
Analytical expressions exist only for the isolated point lens~\cite{peters1974index} and series expansions have been developed for a few symmetric lenses~\cite{Matsunaga:2006uc, Wright_2022}, but even these solutions become costly to evaluate at high frequency $\times$ lens mass. Waveforms for high-mass lensing objects could allow us especially to probe the large-scale structure of the universe using LISA ~\cite{Gao:2021sxw,Caliskan:2022hbu,Savastano:2023spl,Caliskan:2023zqm, Brando:2024inp}. 

Several publicly available packages have been developed for GW lensing. 
LensingGW\footnote{\url{https://gitlab.com/gpagano/lensinggw}} \cite{Pagano:2020rwj} is restricted to GO, and thus large lens masses. \textsc{gravelamps}\footnote{\url{https://git.ligo.org/mick.wright/Gravelamps}} \cite{Wright_2022} uses a combination of numerical integration and series expansion for symmetric lenses. 
\texttt{Glworia}\footnote{\url{https://github.com/mhycheung/glworia}} \cite{cheung2024probingminihalolensesdiffracted} employs contour integration to compute amplification factors for symmetric lenses. 
However, these codes are not yet fast enough for sampling over large sets of parameters: applications to parameter estimation require pre-computing and interpolating the lensing diffraction effects.
Dense distributions of stars embedded in galactic-scale lenses have been analyzed, but all analyses have relied on private codes~\cite{Diego:2019lcd,Mishra:2021xzz,Cheung:2020okf,Shan:2022xfx,Shan:2023ngi,Shan:2023qvd,mishrapopbiases}.

In this work, we present a series of fast and accurate algorithms to compute wave-optics lensing diffraction and their implementation into \textit{Gravitational Lensing of Waves} (\glow{}), a flexible and modular software package in Python \& C. 
The key features of the package are:
\begin{itemize}
    \item Wave-optics lensing methods for general lenses and configurations.
    \item Optimized algorithms for both symmetric lenses and the single-image regime.
    \item Bespoke treatment of singular contributions and lens-plane asymptotics.
    \item Multiple efficient methods for Fourier transform.
    \item A catalog of commonly used lens profiles and flexibility to add any arbitrary lens to the code.
    \item A fast and accurate implementation of the point lens and
    the singular isothermal sphere.
    \item Tunable precision parameters to control accuracy and speed.
    \item Computation of lensed waveforms, in the time and frequency domains.
\end{itemize}
\glow{} is publicly available
\href{https://github.com/miguelzuma/GLoW_public}{\faGithub}
\footnote{\url{https://github.com/miguelzuma/GLoW_public}}, for updated details see the documentation 
\href{https://miguelzuma.github.io/GLoW_public/index.html}{\small\faBook}
\footnote{\url{https://miguelzuma.github.io/GLoW_public/index.html}}. 
Please acknowledge its use by citing this paper. The scripts used to generate the figures in
this paper are also available in the public repository. We point the reader to Refs. \cite{Savastano:2023spl,Brando:2024inp,Zumalacarregui:2024ocb} for some of the GW lensing science case studies that have been enabled by \glow{}.

The rest of this article is organised as follows, Sec.~\ref{sec:form} introduces the concepts and equations involved in gravitational lensing computations in the wave-optics regime. 
Sec.~\ref{sec:time} describes the algorithms used for the time-domain computation, 
Sec.~\ref{sec:freq} describes the regularization and conversion to the frequency domain. 
Sec.~\ref{sec:analytic} presents efficient analytical results in some symmetric lenses. 
Sec.~\ref{sec:code} presents the structure of the code and discusses its performance.
We provide our conclusions in Sec.~\ref{sec:concl}. The appendices present a catalogue of implemented lenses (\ref{app:lenses}), expressions for the regularization functions (\ref{app:reg}), and tests of 
\glow{}'s precision (\ref{sec:code_prec}).

\section{Lensing formalism}\label{sec:form}

In this Section we will describe the basic setup of gravitational lensing
\cite{Schneider1999-tk} and notation followed by the code. We work in units where $c = 1$.

    The lensing of gravitational waves is characterized by the \textit{amplification factor} $F(f)$ multiplying an unlensed strain $\tilde{h}_0$ in the frequency domain as follows:
    \begin{equation}
    \label{F_def}
        F(f) \equiv \frac{\tilde{h}(f)}{\tilde{h}_0(f)}  \ .
    \end{equation}
    Here, $\tilde{h}_0(f)$ and $\tilde{h}(f)$ are Fourier transforms of the unlensed and lensed strains respectively. 
    For a lens located at redshift $z_{L}$, the distance $D_{LS}$ between the lens and the source is
    \begin{align}
        D_{LS} &= D_S - \frac{1+z_L}{1+z_S}D_L \ ,
    \end{align}
    where $D_S$ and $D_L$ are the angular diameter distances to the source and the lens respectively. We can then define an effective distance 
    \begin{equation}
        d_\text{eff} \equiv \frac{D_LD_{LS}}{(1+z_L)D_S}\ .
    \end{equation}
    For positions on the lens plane $\v{\xi}$ and on the source plane $\v{\eta}$, we define the dimensionless parameters
    \begin{equation}
        \v{x} = \frac{\v{\xi}}{\xi_0}\,,\qquad \v{y}=\frac{D_L}{\xi_0D_S}\v{\eta}\ .
    \end{equation}
    Here, $\v{y}$ is known as the \textit{impact parameter} and $\xi_0$ is a dimensionful (but otherwise arbitrary)
    scale, typically chosen depending on the lens. See Appendix \ref{app:lenses} for more details on the
    choice of $\xi_0$. 
    
    For a given density profile $\rho(\v{r})$ of a lens, the projected mass density is obtained by integrating in the $z$-direction perpendicular to the lens plane:
    \begin{equation}\label{eq:def_Sigma}
        \Sigma(\v{\xi}) = \int^\infty_{-\infty}\di z\ \rho(\v{\xi}, z)\ .
    \end{equation}
    The lensing potential $\psi(\boldsymbol{x})$ for the particular lens is then found by solving
    \begin{equation}
    \label{eq:nabla_psi}
        \nabla^2_{\v{x}}\psi(\boldsymbol{x})=2\kappa(\boldsymbol{x}) \ ,
    \end{equation}
    where the \textit{convergence} $\kappa$ is given by
    \begin{equation}
        \kappa(\v{x}) = \frac{\Sigma(\xi_0\v{x})}{\Sigma_\text{cr}}\,,
    \end{equation}
    and the \textit{critical density} $\Sigma_\text{cr}$ is
    \begin{equation}
        \Sigma_\text{cr} = \frac{1}{4\pi G(1+z_L)d_\text{eff}}\ .
    \end{equation}
    Here, $G$ is Newton's constant. 
    We can solve \eqref{eq:nabla_psi} for the lensing potential using the 2D Green's function
    \begin{equation}
        \psi(\v{x}) = \frac{1}{\pi}\int\di^2x'\,\kappa(\v{x}')\log|\v{x}-\v{x}'|\ .
    \end{equation} 

    The lensing potential is then incorporated into the amplification factor via the \textit{Fermat potential}, which is defined as
    \begin{align}
        \phi(\v{x}, y) &= \frac{1}{2}|\v{x}-\v{y}|^2 - \psi(\v{x})\\
                       &= \frac{1}{2}(x_1^2+x_2^2+y^2) - x_1y - \psi(x_1, x_2)\ ,\\
        \tilde{\phi} &\equiv \phi - t_\text{min}\ ,
    \end{align}
    where $t_\text{min}$ is the minimum of the Fermat potential, such that 
    $\tilde{\phi}(\v{x}_\text{min})=0$. We also choose the orientation of the axes such
    that the $x_1$ axis is aligned with the impact parameter $\v{y}$.
    The amplification factor \eqref{F_def} is defined in terms of the Fermat potential as
    \begin{equation}\label{eq:Fw_def}
        F(w) = \frac{w}{2\pi \ii}\int\di^2 x\exp\Big(\ii w\tilde{\phi}(\v{x}, y)\Big)\ .
    \end{equation}
    Here we introduced the \textit{dimensionless frequency}
    \begin{equation}
        w \equiv 8\pi GM_{Lz}f\ ,
    \end{equation}
    where $M_{Lz}$ is the redshifted \textit{effective lens mass}
    \begin{equation}\label{eq:def_MLz}
        M_{Lz} \equiv \frac{\xi_0^2}{4Gd_\text{eff}}\ .
    \end{equation}
    Defining the Fourier transform as
    \begin{align}
        \mathcal{F}\big[f(x)\big] &\equiv \int^{\infty}_{-\infty}\di x\,
            \ee^{-\ii\omega x}f(x)\ ,
    \end{align}
    the time-domain version of the amplification factor can be written as
    \begin{align}
        F(w) &= \frac{w}{2\pi\ii}\int^\infty_{-\infty}\di\tau\,\ee^{\ii w\tau}I(\tau)
            =\frac{w}{2\pi\ii} \mathcal{F}^{*}\big[I(\tau)\big]\ ,\\
        I(\tau) &= \int\di^2x\, \delta\Big(\tilde{\phi}(\v{x})-\tau\Big)\ ,
    \end{align}
    where $^{*}$ denotes complex conjugation. In many cases, we will use the notation
    \begin{align}\label{eq:It_def}
        I(t) \equiv \int\di^2x\, \delta\Big(\phi(\v{x})-t\Big)\ ,
    \end{align}
    where $t\equiv \tau + t_\text{min}$. The conversion to physical time and frequency
    is
    \begin{subequations}
    \begin{align}
        t_\text{phys} &= \frac{\xi_0^2}{d_\text{eff}}t = 4GM_{Lz}t\,,\\
        f &= \frac{1}{2\pi}\frac{d_\text{eff}}{\xi_0^2}w = \frac{w}{8\pi GM_{Lz}}\ .
    \end{align}
    \end{subequations}    

\section{Time domain}\label{sec:time}
In this Section, we provide an overview of the algorithms used to compute the time-domain version of the amplification factor $I(\tau)$.
    \subsection{Contour method}\label{sec:contour}
\begin{figure*}[ht]
            \includegraphics[scale=0.8]{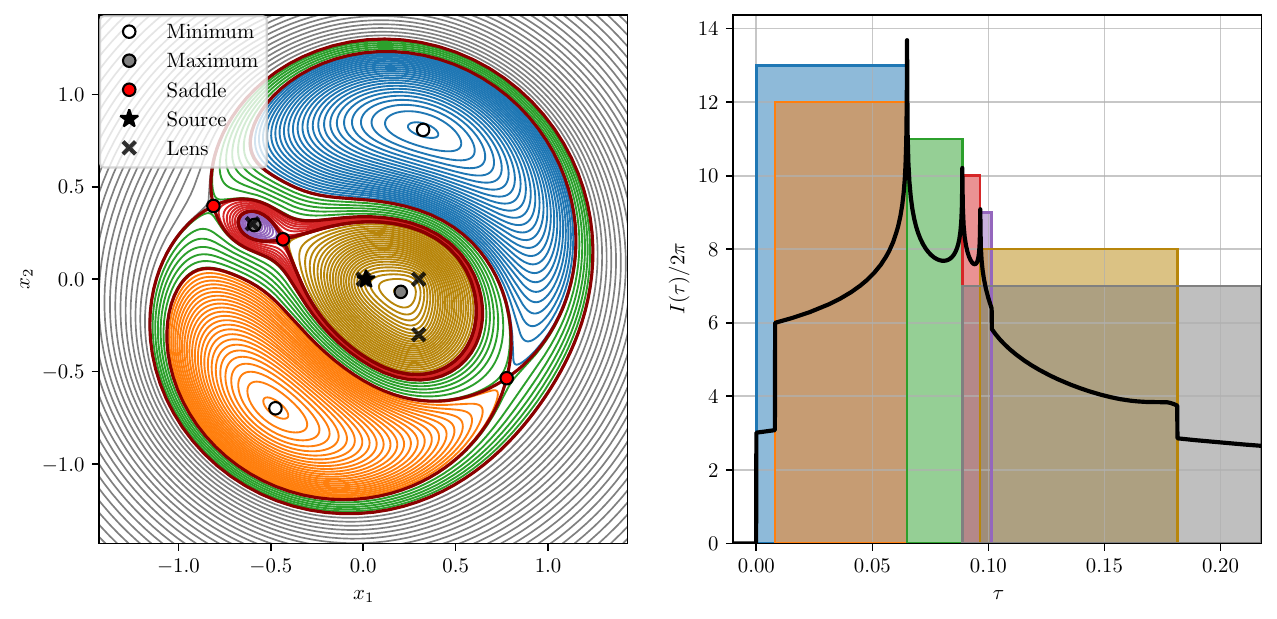}
            \caption{Non-trivial application of the contour method in Sec. \ref{sec:contour}. 
            The lens is composed of four CISs (see App. \ref{app:lenses} for the definition). 
            \textbf{Left:} Equal time delay contours and critical points of the Fermat potential.
            Each family of contours starts and ends at a critical point, and is represented with 
            a different color. The outermost family of contours extends to infinity. 
            \textbf{Right:} Each contour, with time delay $\tau$, contributes to a point in 
            $I(\tau)$. The range where each family contributes is represented with colored 
            boxes. The boxes are displayed with different heights to aid visualization.}
            \label{fig:simple_contours}
        \end{figure*}

        The original proposal of computing $I(\tau)$ as a contour integral dates back to
        the work of Ulmer and Goodman in 1994 \cite{Ulmer:1994ij}. Since then, several
        works have applied this method \cite{Tambalo:2022plm, Tambalo:2022wlm, Savastano:2023spl, Cheung:2024ugg}. 
        We will assume for now that there is a single image, i.e.~the global minimum. We 
        can change from Cartesian to polar coordinates centered at this minimum
        \begin{align}\label{eq:x1x2_def}
            x_1 &= x_1^0 + R\cos\theta\ ,\\
            x_2 &= x_2^0 + R\sin\theta\ ,
        \end{align}
        such that $\phi(x_1^0, x_2^0)=t_\text{min}$. With this change of coordinates, the
        time-domain integral can be rewritten as
        \begin{equation}
            I(t) = \int R\,\di R\,\di\theta\,\delta\Big(\phi(R, \theta)-t\Big)\ .
        \end{equation}
        If $\partial_R\phi\neq 0$, we can invert
        \begin{equation}
            \phi(R, \theta) = t\ ,
        \end{equation}
        to obtain $R(\theta, t)$. We can then solve the integral over the $\delta$
        function, plugging in this solution
        \begin{equation}
            I(t) = \int^{2\pi}_0\frac{R(\theta, t)}{|\partial_R\phi|}\di\theta\ .
        \end{equation}
        Finally, the system of differential equations that must be solved to find both
        the curve $R(\theta, t)$ and $I(t)$ is
        \begin{align}\label{eq:sys_ODE_def_nonparam}
            \d{I}{\theta} &= \frac{R}{|\partial_R\phi|}\ ,\\
            \d{R}{\theta} &= -\frac{\partial_\theta\phi}{\partial_R\phi}\ .
        \end{align}
        Since we are just interested in $I(t)$, we can integrate this system from 
        $\theta=0$ to $2\pi$, with initial conditions $I(\theta=0, t)=0$ and 
        $R(\theta=0, t)$ such that $\phi(R(0, t), 0)=t$. The previous condition 
        $\partial_R\phi\neq 0$ can be violated when the lensing effects are very aggressive
        and the contours are very deformed. In this case, we must find a parametric 
        representation of the constant time-delay curve as $R(\sigma, t)$, 
        $\theta(\sigma, t)$. We will choose the parameterization 
        \begin{align}
            \d{R}{\sigma} = -\partial_\theta\phi\ ,\quad
            \d{\theta}{\sigma} = \partial_R\phi\ ,
        \end{align}      
        that transforms the integral into
        \begin{equation}
            I(t) = \oint R(\sigma, t)\di\sigma\ .
        \end{equation}
        The problem is then equivalent to solving the following system of differential equations
        \begin{subequations}\label{eq:sys_ODE_def}
        \begin{align}
            \d{I}{\sigma} &= R\ ,\\
            \d{R}{\sigma} &= -\partial_\theta\phi\ ,\\
            \d{\theta}{\sigma} &= \partial_R\phi\ .
        \end{align}
        \end{subequations}
        This time, we must integrate from $\sigma=0$ until we close the curve, i.e. 
        $\theta(\sigma_f, t) = 2\pi$ and $R(\sigma_f, t)=R(0, t)$. The initial conditions 
        are chosen as before. The method can be generalized from the single-image case
        that we have developed here to a generic strong-lensing scenario. It can be 
        summarized as follows:
        \begin{enumerate}
            \item Find all the critical points solving the lens equation $\v{\nabla}\phi=0$.
            \item Some lenses also present special points, like singularities and cusps. 
                These points may introduce discontinuities and divergences in 
                $\v{\nabla}\phi$. However, if we use a regularized version of the lens 
                these points reduce to a standard critical point, so in the following 
                discussion we will also refer to them collectively as ``critical points''. 
            \item Once we have found all the critical points we can divide the lens plane
                into regions covered by different families of contours. The regions are
                separated by critical curves, which are curves of constant time delay that cross
                a saddle point.  
            \item Each of these families of contours starts at a critical point with time
                delay $t^i_0$ and dies at another critical point $t^i_f$. These contours
                then contribute to $I(t)$ in the range $[t^i_0, t_f^i]$. In each of these
                regions we can choose coordinates \eqref{eq:x1x2_def}, with $\v{x}^0$
                being the critical point at the center of the contours, and integrate
                the system \eqref{eq:sys_ODE_def} to obtain the contribution to $I(t)$.
            \item Finally, once we have identified all the families of contours and their
                contributions, we can add them to find the total $I(t)$.
        \end{enumerate}
        Notice that, after finding the regions in the lens plane, the integration over
        each contour is independent, so the algorithm can be trivially parallelized. 
        A non-trivial example of this method is shown in Fig. \ref{fig:simple_contours}.
        
    \subsection{Integral for symmetric lenses}\label{sec:single_integral}
        In the axisymmetric case, i.e. $\psi(\v{x}) = \psi(x)$, it is possible to 
        reduce the problem in the time domain to the computation of an ordinary
        integral in one variable. First, we must choose coordinates adapted to the 
        symmetry of the lensing potential
        \begin{subequations}
        \begin{align}
            x_1 &= r\cos\varphi\ ,\\
            x_2 &= r\sin\varphi\ .
        \end{align}
        \end{subequations}
        Defining $z\equiv \cos\varphi$, we can rewrite \eqref{eq:It_def} as
        \begin{align}
            \phi(r, z) &= \frac{1}{2}(r^2+y^2) - yrz - \psi(r)\ ,\\
            I(t) &= \int^{1}_{-1}\frac{\di z}{\sqrt{1-z^2}}\int^\infty_0 r\,\di r\,
                \delta\Big(\phi(r, z) - t\Big)\ .
        \end{align}
        Solving the integral over $z$ we get
        \begin{align}
            I(t) &= \int^\infty_0\frac{2\di r}{y\sqrt{1-z_*^2}}\Theta(1-z_*^2)\ ,\nonumber\\
                z_* &\equiv \frac{1}{2yr}\left(r^2-2\psi(r)-t\right)\ ,
        \end{align}
        where $\Theta$ is the Heaviside step function. The computation can be simplified
        further if we split the integrand into the regions where it is non-zero, i.e.
        $r\in(r_\text{min}, r_\text{max})\rightarrow |z_*|<1$,
        \begin{align}
            I(t) &= \sum_i \int^{r^i_\text{max}}_{r^i_\text{min}}\alpha(r) 
                \,\di r\ ,\\
            \alpha(r) &\equiv \frac{2r}{\sqrt{-\phi_+\phi_-}}\ ,\\
            \phi_\pm &\equiv \frac{1}{2}r^2 + \frac{1}{2}y^2 - \psi(r) - t \mp ry\ .\label{eq:phi_pm_def}
        \end{align}
        At the limits $r_\text{max}$ and $r_\text{min}$ we have $\phi_+\phi_-=0$, so even
        though the function is integrable, it can be hard to compute numerically. We make
        one more change of variables that smooths the integrand
        \begin{equation}
            \zeta = \begin{cases}\displaystyle
                \sqrt{\frac{r-r_\text{min}}{r_\text{mid}-r_\text{min}}}\ , & 
                    r\in (r_\text{min}, r_\text{mid})\\[10pt] \displaystyle
                \sqrt{\frac{r_\text{max}-r}{r_\text{max}-r_\text{mid}}}\ , & 
                    r\in (r_\text{mid}, r_\text{max})
            \end{cases}
        \end{equation}
        where $r_\text{mid}\equiv (r_\text{max}+r_\text{min})/2$. Using this new variable
        we just need to compute a single integral
        \begin{align}\label{eq:singleintegral_def}
            I(t) = 2\int^1_0\zeta\,\di\zeta\sum_i \sqrt{\Delta^i}\Big\{
                &\alpha(r^i_\text{max}-\Delta^i\zeta^2)\nonumber\\
                &+\alpha(r^i_\text{min}+\Delta^i\zeta^2)\Big\}\ ,
        \end{align}
        with $\Delta^i\equiv (r_\text{max}^i-r_\text{min}^i)/2$.
        The whole algorithm can be summarized as follows:
        \begin{enumerate}
             \item Find all the critical points (and special points, see Sec. \ref{sec:contour}).
                 In the axisymmetric case, all the critical points lie in the $x_1$ line 
                 (i.e. $x_2=0$) so now we just need to deal with a much simpler
                 1D root-finding problem. 
             \item For each $t$, determine the regions where $\phi_+(r)\phi_-(r)<0$ \eqref{eq:phi_pm_def} 
                 by finding the values $(r^i_\text{min}, r^i_\text{max})$ where 
                 $\phi_+\phi_-=0$.
             \item Perform the integral \eqref{eq:singleintegral_def}, summing over all the
                 $i$ regions where $\phi_+\phi_-<0$.
        \end{enumerate}
        
    \subsection{Grid method}\label{sec:grid}
        Finally, another way to solve the time-domain integral, first proposed in 
        \cite{Diego:2019lcd}, is to compute it directly as a 
        surface integral. This method has been applied in a number of works
        \cite{Diego:2019rzc, Cheung:2020okf, Yeung:2021chy, Shan:2022xfx}. Starting again
        with the time-domain version of the amplification factor
        \begin{equation}
            I(t) = \int\di^2x\,\delta(\phi(\v{x})-t)\ ,
        \end{equation}
        the simplest approach we can follow to compute this integral is to represent the delta 
        function as
        \begin{equation}
            \delta_n(x) = \left\{\begin{array}{ll}
                0\ , &x<-1/2n\\
                n\ , &-1/2n<x<1/2n\\
                0\ , &x> 1/2n
            \end{array}\right.
        \end{equation}
        when $n\to \infty$. In this way, we obtain a discrete representation
        \begin{equation}
            I(t) \simeq I_i\ ,\quad \text{for}\  t\in [t_i-\Delta t_i/2, t_i+\Delta t_i/2]\ ,
        \end{equation}
        that converges to the real result as we reduce the size of the boxes $\Delta t_i$.
        The approach we follow to implement this method is the following:
        \begin{enumerate}
            \item Find the global minimum of the Fermat potential, $t_\text{min}$. 
            \item Create a temporal grid from $t_\text{min}$ to a given $t_\text{max}$, 
                logarithmically spaced.
            \item Define a spatial grid and evaluate $\phi(\v{x})$ on this grid.
            \item Build the discrete representation $I_i$ as a histogram. 
        \end{enumerate}

\section{Frequency domain}\label{sec:freq}
Having computed the time-domain integral $I(\tau)$, in this Section, we detail the regularization and Fourier transform procedures we follow to transform $I(\tau)$ into the frequency domain.
    \subsection{Regularization}\label{sec:regularization}
        Once the time-domain integral has been computed, we must compute its Fourier 
        transform (FT) to obtain the amplification factor
        \begin{equation}
            F(w) = \frac{w}{2\pi\ii}\int^\infty_{-\infty}\di\tau\,\ee^{\ii w\tau}I(\tau)
                 =\frac{w}{2\pi\ii} \mathcal{F}^{*}\big[I(\tau)\big]\ .
        \end{equation}
        However, the presence of discontinuities and singularities in $I(\tau)$ makes this
        seemingly-simple operation much more complicated in practice. In
        \cite{Tambalo:2022plm}, a regularization approach was proposed to overcome 
        this problem. The main idea isto split the result into a regular and singular 
        part, 
        \begin{subequations}
        \begin{align}
            I(\tau) &= I_\text{reg}(\tau) + I_\text{sing}(\tau)\ ,\\
            F(w) &= F_\text{reg}(w) + F_\text{sing}(w)\ ,
        \end{align}
        \end{subequations}
        such that $I_\text{sing}$ is an analytical expression that we can Fourier transform
        analytically (i.e. $F_\text{sing}$ is also known) and $I_\text{reg}$ is easier to
        Fourier transform numerically. The regularization scheme proposed in 
        \cite{Tambalo:2022plm} was to add for each minimum (type I image)
        \begin{subequations}\label{eq:old_reg_I}
        \begin{align}
            I^\text{\tiny m}_\text{\tiny sing}(\tau) &\equiv 2\pi \sqrt{\mu_j}\,\Theta(\tau-\tau_j)\ ,\\
            F^\text{\tiny m}_\text{\tiny sing}(w) &= \sqrt{\mu_j}\,\ee^{\ii w\tau_j}\ ,
        \end{align}
        \end{subequations}
        for each maximum (type II image)
        \begin{subequations}\label{eq:old_reg_II}
        \begin{align}
            I^\text{\tiny M}_\text{\tiny sing}(\tau) &\equiv 2\pi \sqrt{\mu_j}\,\Theta(\tau_j-\tau)\ ,\\
            F^\text{\tiny M}_\text{\tiny sing}(w) &= -\sqrt{\mu_j}\,\ee^{\ii w\tau_j}\ ,
        \end{align}
        \end{subequations}
        and for each saddle point (type III image)
        \begin{subequations}\label{eq:old_reg_III}
        \begin{align}
            I^\text{\tiny s}_\text{\tiny sing}(\tau) &\equiv -2\sqrt{\mu_j}\ee^{-|\tau-\tau_j|/T}\log|\tau-\tau_j|\ ,\\
            F^\text{\tiny s}_\text{\tiny sing}(w) 
                &= \frac{2\ii w}{\pi}\sqrt{\mu_j}\ee^{\ii w\tau_j}\ \Re(\mathcal{I})\ ,\\
                    \mathcal{I} &\equiv \int^\infty_0\di t\log(t)\ee^{-t/T+\ii wt} \nonumber\\
                &= -\frac{\gamma_E + \log(T^{-1}-\ii w)}{T^{-1}-\ii w}\ ,
        \end{align}
        \end{subequations}
        where $T$ is a free parameter. 
        
        This scheme removes all the discontinuities and singularities associated with
        the geometric optics result, typically reducing the error in the FT at high
        frequencies. This regularization has the clear advantage of being very easy to
        evaluate in terms of elementary functions but also presents some drawbacks, 
        e.g. $I_\text{reg}(\tau)$ is not zero for $\tau<0$ and $I_\text{sing}(\tau)$ 
        contains free parameters. 
        
        The most important problem, however, is the presence of errors in the FT at low
        frequencies. These arise from the high-$\tau$ behaviour of $I(\tau)$, usually a
        power-law tail. The new regularization scheme that we propose in this work can 
        handle both the geometric-optics singularities and these long tails, at the cost
        of introducing more complex regularizing functions. We find that, numerically, 
        the tradeoff of reducing the complexity of the FT while increasing the cost
        of computing $I_\text{sing}$ and $F_\text{sing}$ is highly beneficial. 
        
        We will assume that the asymptotic behaviour of $I(\tau)$ is
        \begin{equation}
            \frac{I(\tau\to\infty)}{2\pi}\sim 1 + \frac{I_\text{asymp}}{\tau^\sigma}\ ,
        \end{equation}
        and define
        \begin{equation}
            \mathcal{C}_\text{M}\equiv \sum_\text{max}\sqrt{|\mu_j|}\ ,\qquad
            \mathcal{C}_\text{m}\equiv \sum_\text{loc min}\sqrt{|\mu_j|}\ ,
        \end{equation}
        where the second sum spans only the local minima, and we will also denote as
        $\sqrt{\mu_\text{min}}$ the magnification of the global minimum. With these 
        definitions we can write the singular contribution
        \begin{align}\label{eq:It_sing}
            \frac{I_\text{sing}(\tau)}{2\pi} &= 
                    \sum_\text{loc min}\sqrt{|\mu_j|}\Theta(\tau-\tau_j)
                    + \sum_\text{max}\sqrt{|\mu_j|}\Theta(\tau_j-\tau)\nonumber\\
                &\quad - \mathcal{C}_\text{M} 
                    + \left(1-\mathcal{C}_\text{m}+\mathcal{C}_\text{M}\right)\Theta(\tau)\nonumber\\
                &\quad +R_0\left(\sqrt{|\mu_\text{min}|}-1-\mathcal{C}_\text{M}+\mathcal{C}_\text{m}, I_\text{asymp}, \sigma;\ \tau\right)\nonumber\\
                &\quad +\sum_\text{saddle}S_\text{full}\left(\frac{2\sqrt{|\mu_j|}}{\pi\tau_j}, \tau_j;\ \tau\right)\ ,
        \end{align}
        and its Fourier counterpart as
        \begin{align}\label{eq:Fw_sing}
            F_\text{sing}(w) &= \sum_\text{loc min}\sqrt{|\mu_j|}\ee^{\ii w\tau_j}
                - \sum_\text{max}\sqrt{|\mu_j|}\ee^{\ii w\tau_j}\nonumber\\
                &\quad +1-\mathcal{C}_\text{m}+\mathcal{C}_\text{M}\nonumber\\
                &\quad +\tilde{R}_0\left(\sqrt{|\mu_\text{min}|}-1-\mathcal{C}_\text{M}+\mathcal{C}_\text{m}, I_\text{asymp}, \sigma;\ w\right)\nonumber\\
                &\quad +\sum_\text{saddle}\tilde{S}_\text{full}\left(\frac{2\sqrt{|\mu_j|}}{\pi\tau_j}, \tau_j;\ w\right)\ .
        \end{align}
        The regularizing functions are lengthy and are collected in Appendix \ref{app:reg}.
        
        \begin{figure*}[ht]            
            \includegraphics[scale=0.8]{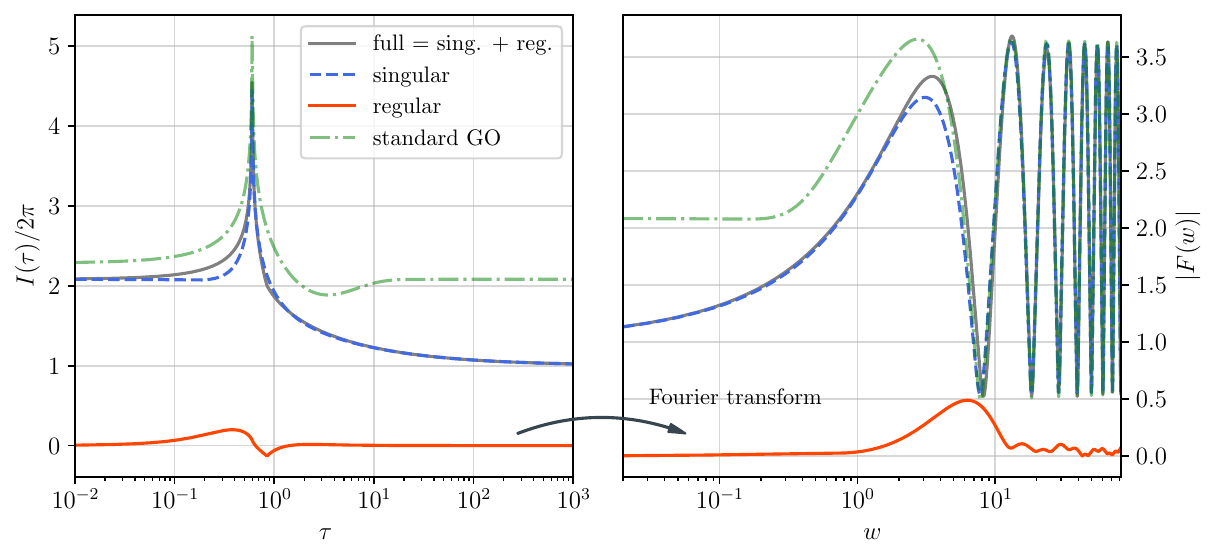}
            \caption{Outline of our regularization procedure. 
            \textbf{Left:} We start by computing the time domain integral, using any of the 
            methods in Sec. \ref{sec:time}. We then regularize it, subtracting the analytical
            singular contribution \eqref{eq:It_sing}. This regularization scheme subtracts the
            saddle point divergence, the step function at $\tau=0$, as well as the slowly 
            decaying tail at large $\tau$. \textbf{Right:} The regular part is then Fourier
            transformed, using any of the methods in Sec. \ref{sec:freq}. Finally, we recover
            the full amplification factor adding back \eqref{eq:Fw_sing}, i.e. the (analytical) 
            Fourier transform of the singular contribution. For comparison, we also show the
            old regularization scheme, standard GO, representing only the singular contribution given by 
            Eqs. \eqref{eq:old_reg_I}-\eqref{eq:old_reg_III}.
            }
            \label{fig:regularization}
        \end{figure*}

    \subsection{FFT with varying resolution}\label{sec:fft_multigrid}
        Choosing the right regularization scheme is an important ingredient to correctly transform
        $I(\tau)$ to the frequency domain, but not the only one. The fastest way to perform the Fourier
        transform is to use the Fast Fourier Transform (FFT), but a naive implementation will face several
        obstacles:
        \begin{itemize}
            \item For typical applications to GW lensing, we will need to compute the amplification factor
                for frequencies spanning several orders of magnitude. Since the FFT samples the functions
                linearly, key features at low frequency (e.g. Fig. \ref{fig:regularization}), will be severely undersampled. 
                The simplest way to overcome this limitation is to extend the Fourier transform to
                higher frequencies than are actually needed, increasing then the number of points and
                effectively improving the sampling at low $w$. This of course has a significant impact
                on the performance.
            \item Even if we do not need denser sampling at lower frequencies, another common problem is 
                the appearance of errors both at high and low frequencies. This can again be mitigated 
                by performing an FFT over a wider range than is actually needed, improving the overall 
                sampling in the time domain. The downside again is a loss of performance. 
            \item Finally, even after computing a larger (slower) FFT, there is another difficulty that
                will slow down the computation. The typical strategy to lens a waveform is to precompute
                $F(w)$ with an FFT and then interpolate it as needed over the waveform frequencies. If we
                compute a very large FFT, to correctly sample the lower frequencies, we will
                oversample the high frequencies, ending up with a grid much larger than actually needed and
                much slower to interpolate. 
        \end{itemize}

        In order to overcome these problems, we generalized a method already considered in 
        \cite{Tambalo:2022plm} and \cite{Tambalo:2022wlm}. The main idea is to construct $F(w)$ out of
        several small FFTs rather than a single, big one. In this way, we sample $I(\tau)$ with varying
        resolution and then we assemble all the contributions. If we want to compute $F(w)$ between two
        frequencies $w_\text{min}$ and $w_\text{max}$, we start by dividing $w$ into 
        logarithmically-spaced frequency ranges $[w^i, 2^N w^i]$, where $N$ is a constant and 
        $w^{i+1} = 2^N w^i$. We then perform independent FFTs in each of this ranges, following the
        same strategy as for the naive FFT: we compute a larger FFT than actually needed and discard the
        low and high frequency range when the errors start creeping in. Finally, we piece all the 
        contributions together. 
        
        This procedure overcomes all the problems outlined before. First, it samples more uniformly the
        amplification factor, without incurring any penalty for going to higher or lower frequencies. 
        Secondly, the performance is much better, since it is much faster to perform several small FFTs 
        rather than a large one. Finally, since the grid is more sparse, the interpolation is also faster.
        The precision achieved with this method is also very good, as shown in Figs. 
        \ref{fig:Fw_SIS_precision} and \ref{fig:Fw_PL_precision}.  
    
    \subsection{Direct Fourier integral}\label{sec:directft}
        The FFT method should always be preferred for applications requiring high speed, since it can 
        compute $F(w)$ automatically over a grid extremely quickly. However, we also wanted to provide an
        alternative method to cross-check the previous scheme. 

        This second approach is only limited by the finite sampling of $I(\tau)$ (and numerical errors) and
        can then be easily improved and tested. The main idea is that, since in most applications we will
        precompute $I(\tau)$ on a grid and linearly interpolate it, the Fourier transform can be computed
        exactly as a sum.
        
        If we represent $I(\tau)$ with a linear-interpolation approximation
        \begin{equation}
            I(\tau)=\begin{cases}
                \displaystyle
                I(\tau=0) + \frac{\tau}{\tau_0}\Big(I(\tau=0) - I_0\Big)\,,
                    \quad &\tau\in [0, \tau_0)\\[8pt]
                \displaystyle
                I_i + \frac{\tau - \tau_i}{\tau_{i+1} - \tau_i}(I_{i+1} - I_i)\,,
                    \quad &\tau\in [\tau_i, \tau_{i+1})\\
                0\,,
                    \quad &\tau\in [\tau_{N-1}, \infty)
            \end{cases}
        \end{equation}
        we can compute the amplification factor analytically and express it as
        \begin{equation}
            F(w) = \sum^{N-2}_{i=-1}J_i\left\{\left(I_{i+1} 
                + \frac{\ii}{w}\frac{\Delta I_i}{\Delta\tau_i}\right)(1-D_i)-\Delta I_i\right\}\ ,
        \end{equation}
        where
        \begin{subequations}
        \begin{alignat}{2}
            \Delta I_i &= I_{i+1} - I_i\,, 
                &\qquad  \Delta I_{-1} &= I_0 - I(\tau=0)\ ,\\
            \Delta\tau_i &= \tau_{i+1}-\tau_i\,, 
                &\qquad  \Delta\tau_{-1} &= \tau_0\ ,\\
            D_i &= \ee^{\ii w\Delta\tau_i}\,,
                &\qquad D_{-1} &= \ee^{\ii w\Delta\tau_{-1}}\ ,\\
            J_i &= J_{i-1}D_{i-1}\,,
                &\qquad J_{-1} &= \frac{1}{2\pi}\ .
        \end{alignat}
        \end{subequations}

        The main drawback of this method is that each frequency must be computed individually.
        Also, one must keep in mind that numerical errors may start creeping in if the sum is 
        extremely large. 

\section{Analytical results}\label{sec:analytic}
Instead of calculating $F(w)$ according to the procedures detailed in Sections \ref{sec:time} and \ref{sec:freq}, some symmetric lenses also have analytic expressions for $F(w)$ which are much more efficient to compute. Here, we provide an overview of the analytic expressions for the amplification factors of two common lens models: point lens and singular isothermal sphere.
    \subsection{Point lens}\label{sec:Fw_pl_analytic}
        The amplification factor \eqref{eq:Fw_def} can be computed analytically for the
        point-mass lens \cite{Takahashi:2003ix}
        \begin{equation}\label{eq:Fw_pl_analytic}
            F_\text{\tiny PL}(w) = u^{\ii u}\ee^{(\pi /2 - 2\ii t_\text{min})u}
                \Gamma(1-\ii u)\,\tensor[_1]{F}{_1}(\ii u, 1, \ii u y^2)\ ,
        \end{equation}
        where $u=w/2$ and $t_\text{min}=(x_\text{min}-y)^2/2-\log(x_\text{min})$ and
        $x_\text{min}=\left(y+\sqrt{y^2+4}\right)/2$. The numerical evaluation of the
        hypergeometric function $\tensor[_1]{F}{_1}(a, b, z)$ with complex parameters is costly and
        it usually requires the use of arbitrary-precision software. However, for the
        particular case $\tensor[_1]{F}{_1}(z, 1, cz)$, we managed to implement a very
        fast version based on four different approximations:
        \begin{enumerate}[I)]
            \item \emph{Large c}. We use the asymptotic formula in chapter 27 of 
                \citep{Temme2014-cl}. For our precision requirements, we compute it
                to third order.
            \item \emph{Small c, large z}. Standard asymptotic formula for large argument
                \cite[\href{http://dlmf.nist.gov/13.7.E2}{(13.7.2)}]{NIST:DLMF}.
            \item \emph{Small c, intermediate z}. We first compute the hypergeometric 
                function $\tensor[_1]{F}{_1}(z, b, cz)$ for sufficiently large $b>1$ using
                the series expansion \cite[\href{http://dlmf.nist.gov/13.2.E2}{(13.2.2)}]{NIST:DLMF}
                and then apply the recurrence relation \cite[\href{http://dlmf.nist.gov/13.3.E2}{(13.3.2)}]{NIST:DLMF}
                to compute the case $b=1$.
            \item \emph{Small c, small z}. Standard series expansion \cite[\href{http://dlmf.nist.gov/13.2.E2}{(13.2.2)}]{NIST:DLMF}.
        \end{enumerate}
        In our code, we calibrated the switches between different approximations to 
        achieve a given level of precision, comparing with the arbitrary-precision 
        implementation in {\sc arb}\footnote{\url{https://github.com/flintlib/arb}} \cite{Johansson2017arb}. 
        We also provide a slower Python implementation based on 
        \texttt{mpmath}\footnote{\url{https://github.com/mpmath/mpmath}} \cite{mpmath}. 
        Fig. \ref{fig:PL_approx} shows the level of agreement between both implementations, highlighting 
        the four regions in parameter space where the previous approximations have been used.
        
    \subsection{Singular isothermal sphere}
        The lensing potential for the singular isothermal sphere (SIS) is
        \begin{equation}
            \psi(\v{x}) = \psi_0|\v{x}|\equiv \psi_0 x\ .
        \end{equation}
        This lens is simple enough that all the geometric optics properties can be
        computed analytically. The minimum of the Fermat potential is located at 
        $\v{x}=(y+\psi_0, 0)$ with time delay and magnification given by
        \begin{align}
            t_\text{min}   &= \frac{1}{2}\psi_0(2y+\psi_0)\ ,\\
            \mu_\text{min} &= 1+\frac{\psi_0}{y}\ .
        \end{align}
        If $y<\psi_0$, there is a second critical point, a saddle point, at $\v{x}=(y-\psi_0, 0)$. The time delay and magnification at the saddle point are
        \begin{align}
            t_\text{saddle}   &= \frac{1}{2}\psi_0(2y-\psi_0)\ ,\\
            \mu_\text{saddle} &= \left| 1-\frac{\psi_0}{y}\right|\ .
        \end{align}
        In addition to this, there is a cusp at the origin.
        It is also possible to compute the full time-domain amplification factor 
        analytically, as we did for the first time in \cite{Savastano:2023spl}. Instead of 
        $\psi_0$ and $y$, we will express the integral as a function of two new 
        variables, $u$ and $R$,
        \begin{equation}\label{eq:def_u_R}
            u \equiv \frac{\sqrt{2\tau}}{\psi_0 + y}\ ,\hspace{1cm}
            R \equiv \frac{\psi_0-y}{\psi_0 + y}\ .
        \end{equation}
        The variable $u$ is a redefined time parameter while $R$ is a constant, 
        ranging between $-1$ and $1$.
        The final result can be compactly expressed as 
        \begin{align}\label{eq:I_sis_analytic}
            I_\text{\tiny SIS}(\tau) 
            &= 
            \frac{8(b-c)}{\sqrt{(a-c)(b-d)}}
            \left[
                \Pi\left(\frac{a-b}{a-c}, r\right) + \frac{c\,K(r)}{(b-c)}
            \right]
            \ ,
        \end{align}
        with
        \begin{equation}
            r \equiv \sqrt{\frac{(a-b)(c-d)}{(a-c)(b-d)}}\ ,
        \end{equation}
        and where $\Pi$ and $K$ are, respectively, the complete elliptic integrals 
        of the third and first kind, see e.g. \cite{gradshteyn2007table}.
        The coefficients $a$, $b$, $c$ and $d$ are functions of the variables 
        $u$ and $R$ defined in \eqref{eq:def_u_R} above. We must however distinguish 
        between three regions
        \begin{itemize}
            \item \emph{Region 1}: $(u>1)$
                \begin{alignat*}{3}
                    a &= 1+u \ ,                   &\qquad  c&= 1-u \ , \\
                    b &= R+\sqrt{u^2 + R^2 -1} \ , &\qquad  d&= R-\sqrt{u^2 + R^2 -1} \ .
                \end{alignat*}
            \item \emph{Region 2}: $(\sqrt{1-R^2}<u<1)$
                \begin{itemize}
                    \item \emph{Case A:} $(R>0)$
                        \begin{alignat*}{3}
                            a &= 1 \ , &\qquad  c&= \sqrt{1-u^2} \ , \\
                            b &= R \ , &\qquad  d&= -\sqrt{1-u^2} \ .
                        \end{alignat*}
                    \item \emph{Case B:} $(R<0)$
                        \begin{alignat*}{3}
                            a &= 1            \ , &\qquad c&= -\sqrt{1-u^2} \ , \\
                            b &= \sqrt{1-u^2} \ , &\qquad d&= R \ .
                        \end{alignat*}
                \end{itemize}
            \item \emph{Region 3}: $(0<u<\sqrt{1-R^2})$
                \begin{alignat*}{3}
                    a &= 1            \ , &\qquad c&= R \ , \\
                    b &= \sqrt{1-u^2} \ , &\qquad d&= -\sqrt{1-u^2} \ .
                \end{alignat*}
        \end{itemize}
        The amplification factor in the frequency domain can be also reduced to a simple
        form, very well suited for numerical computations. First, we can rewrite it as
        \begin{align}
            F_\text{\tiny SIS}(w) &= \frac{w}{2\pi\ii}\int\di^2x\,\ee^{\ii w\phi(\v{x})}\nonumber\\
                &= \frac{w\,\ee^{\ii w y^2/2}}{2\pi\ii}
                    \int^{\pi}_{-\pi}\di\theta\int^\infty_0r\di r\, 
                    \ee^{\ii w r(r/2 - y\cos\theta -\psi_0)}\ .     
        \end{align}
        The radial integral can be expressed in terms of Fresnel integrals
        \cite[\href{http://dlmf.nist.gov/7.2.iv}{(7.2)}]{NIST:DLMF}, and the final 
        expression is 
        \begin{align}\label{eq:Fw_sis_analytic}
            F_\text{\tiny SIS}(w)= \ee^{\ii wy^2/2}\left\{1 +\int^\pi_0\!\!\di\theta\,
                \alpha(\theta)\Big(f(-\alpha)-\ii\,g(-\alpha)\Big)\right\}\,,
        \end{align}
        where
        \begin{equation}
            \alpha(\theta)\equiv \sqrt{\frac{w}{\pi}}\left(\psi_0+y\cos\theta\right)\ .
        \end{equation}
        We implemented the Fresnel integrals following the simple prescription given in \cite{boersma1960computation},
        which achieve an accuracy of $10^{-8}$, more than enough for our applications. 

\section{Code structure and performance}\label{sec:code}
    \subsection{Outline}
\begin{figure}[ht]
            \includegraphics[scale=0.3]{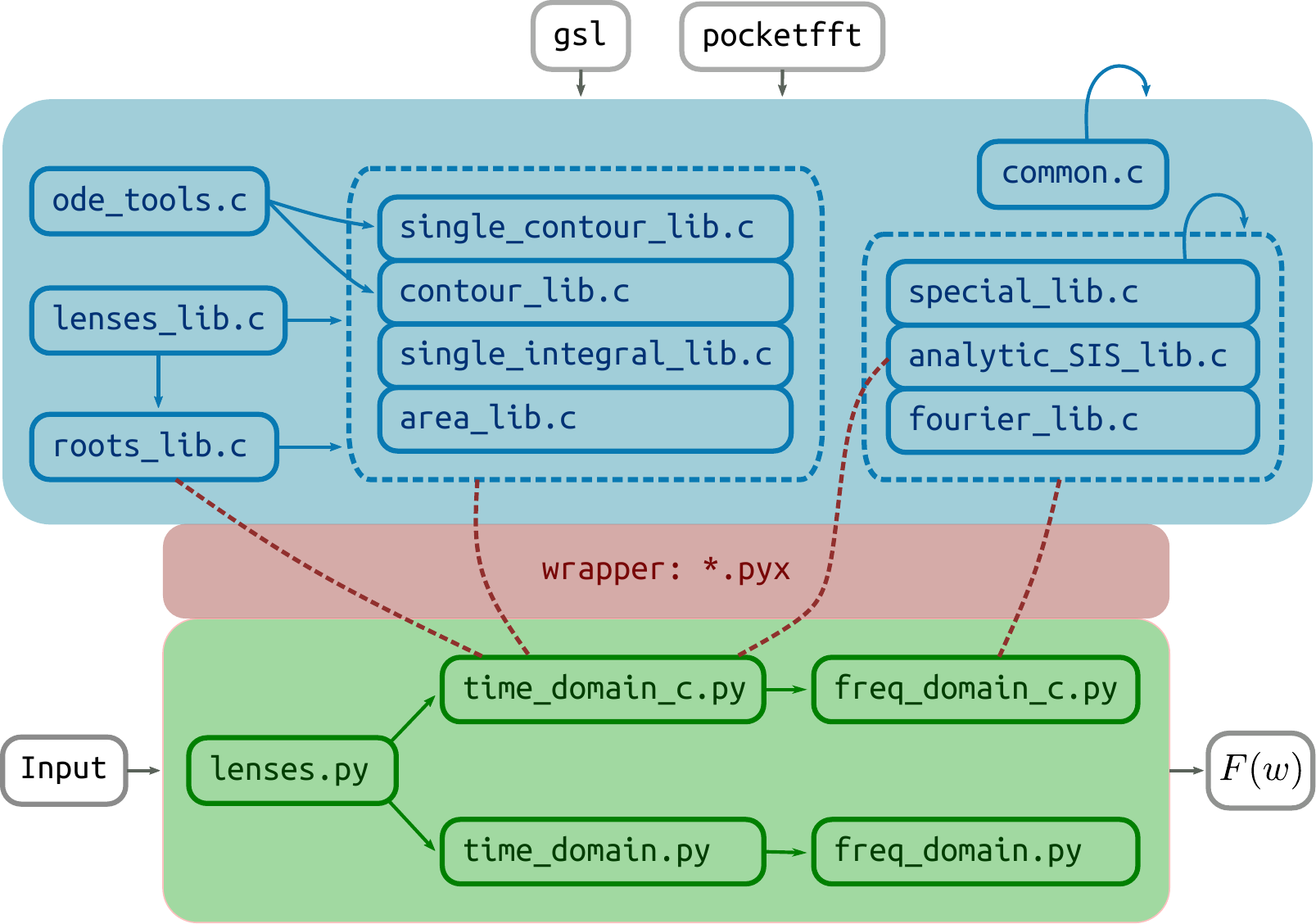}
            \caption{\glow{}'s structure. \glow's core is a standalone C library
            that is interfaced with a Python frontend through a Cython 
            wrapper. Once the library and the wrapper are compiled, the user only needs basic 
            Python knowledge to operate the code. We also provide additional Python utilities 
            to lens waveforms and to transform between physical units and lensing dimensionless 
            units, as well as tutorials and an online documentation.}
            \label{fig:code_diagram}
        \end{figure}

        \glow's main goal is to compute the amplification factor starting from a given lens configuration. 
        Our intention when building \glow{} was
        to create a code that is, first and foremost, fast and supports generic lens configurations.
        To achieve this goal, we developed new methods, explained in the preceding sections, and 
        implemented them in a C library. Furthermore, we wanted to have a modular code that was easy to use 
        by the end user. We have achieved this by developing a Cython wrapper for this library, so that the 
        code can be used entirely from Python. In addition, we also provide a full Python implementation of 
        many of the methods present in the library, although not for all methods, and those that are implemented are not nearly as powerful. Finally, we have a very limited number of external dependencies. The main one is the GNU Scientific 
        Library (GSL)\footnote{\url{https://www.gnu.org/software/gsl/}} \cite{galassi2002gnu}, that can be 
        readily installed from public repositories. We also make use of 
        \texttt{pocketfft}\footnote{\url{https://gitlab.mpcdf.mpg.de/mtr/pocketfft}}, that is included in the 
        source files. The code is organized as shown in Fig. \ref{fig:code_diagram}.
    
        The logic behind the code structure is very simple. There are three main Python modules that must be
        used in succession and create three objects: $\text{Lens}\to I(\tau)\to F(w)$. A simple \glow{}
        session looks like this:
\begin{lstlisting}
import numpy as np
from glow.lenses import Psi_SIS
from glow.time_domain_c import It_SingleIntegral_C
from glow.freq_domain_c import Fw_FFT_C

Psi = Psi_SIS()
It  = It_SingleIntegral_C(Psi, y=0.3)
Fw  = Fw_FFT_C(It)

# computation done, we can evaluate now
ws = np.geomspace(1e-2, 1e2, 1000)
Fws = Fw(ws)
\end{lstlisting}
    
        \begin{figure*}[ht]
            \includegraphics[scale=0.8]{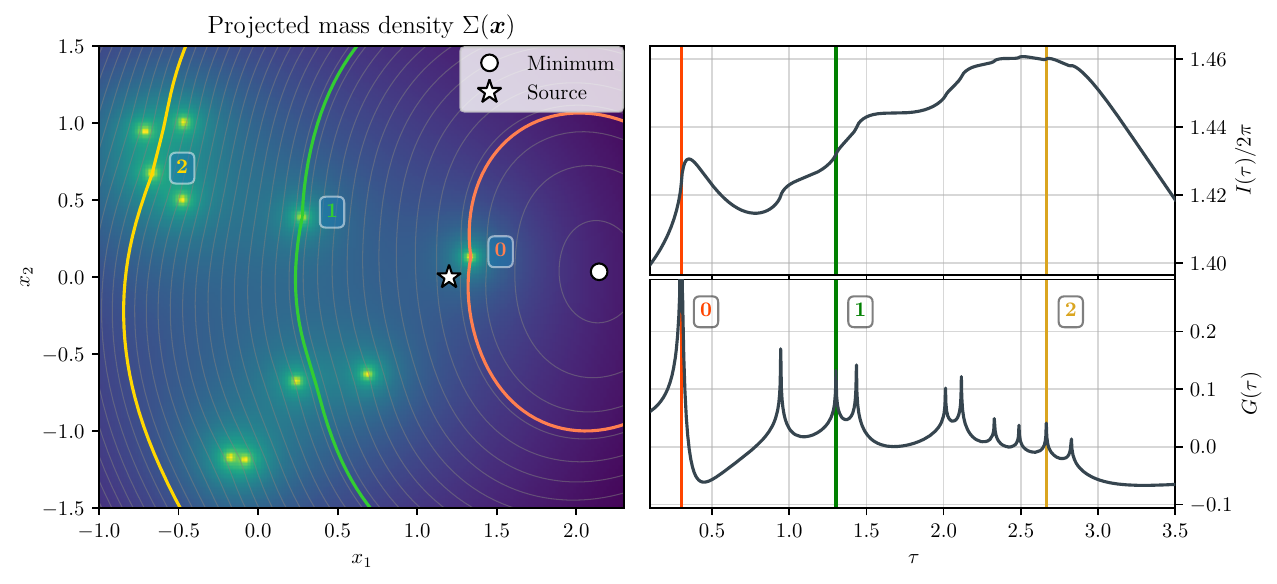}
            \caption{Example of a non-trivial weak lensing scenario, with 10 SISs. The central 
            cusp of the SIS produces small deformations in the time-domain integral. Even more
            dramatic is their effect in the Green function, $G(\tau)\equiv \di I/\di\tau/2\pi$, 
            where each lens produces a distinct peak. We highlighted three of them, together with 
            the contour that passes through the center of the lens.}
            \label{fig:WL_example}
        \end{figure*}

        \begin{figure*}[ht]
            \includegraphics[width=0.99\textwidth]{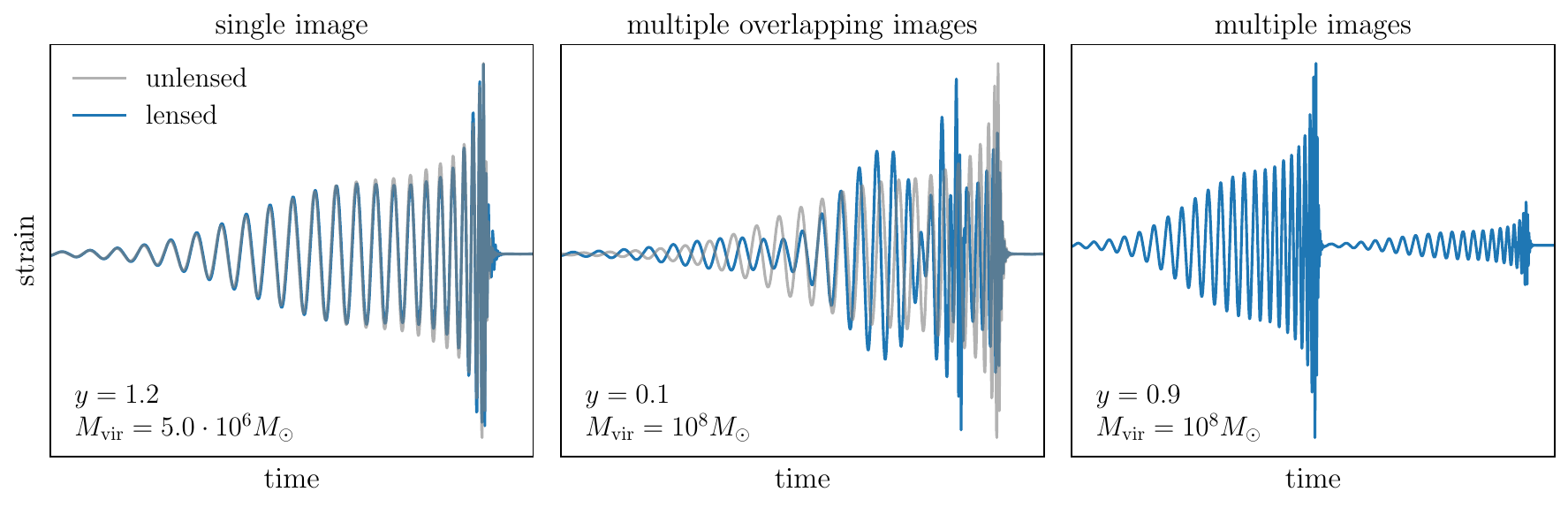}
            \caption{Example of time-domain GW waveforms. Blue curves represent the lensed waveform, while the unlensed signal is plotted in grey. Here, we considered an equal-mass non-spinning BBH with $M_{\rm BBH}=100\, M_\odot$, $z_S=0.3$ and lensed by an SIS lens at $z_L = 0.15$. Notice that all the lensed waveforms are produced using \glow{}'s analytic SIS implementation of the full wave-optics amplification factor. In this regard, we would like to emphasize that we recover 
            remarkably well what is expected in the GO limit \emph{without} using any GO approximation.}
            \label{fig:td_waveform}
        \end{figure*}

        \begin{figure*}[ht]
            \includegraphics[scale=0.38]{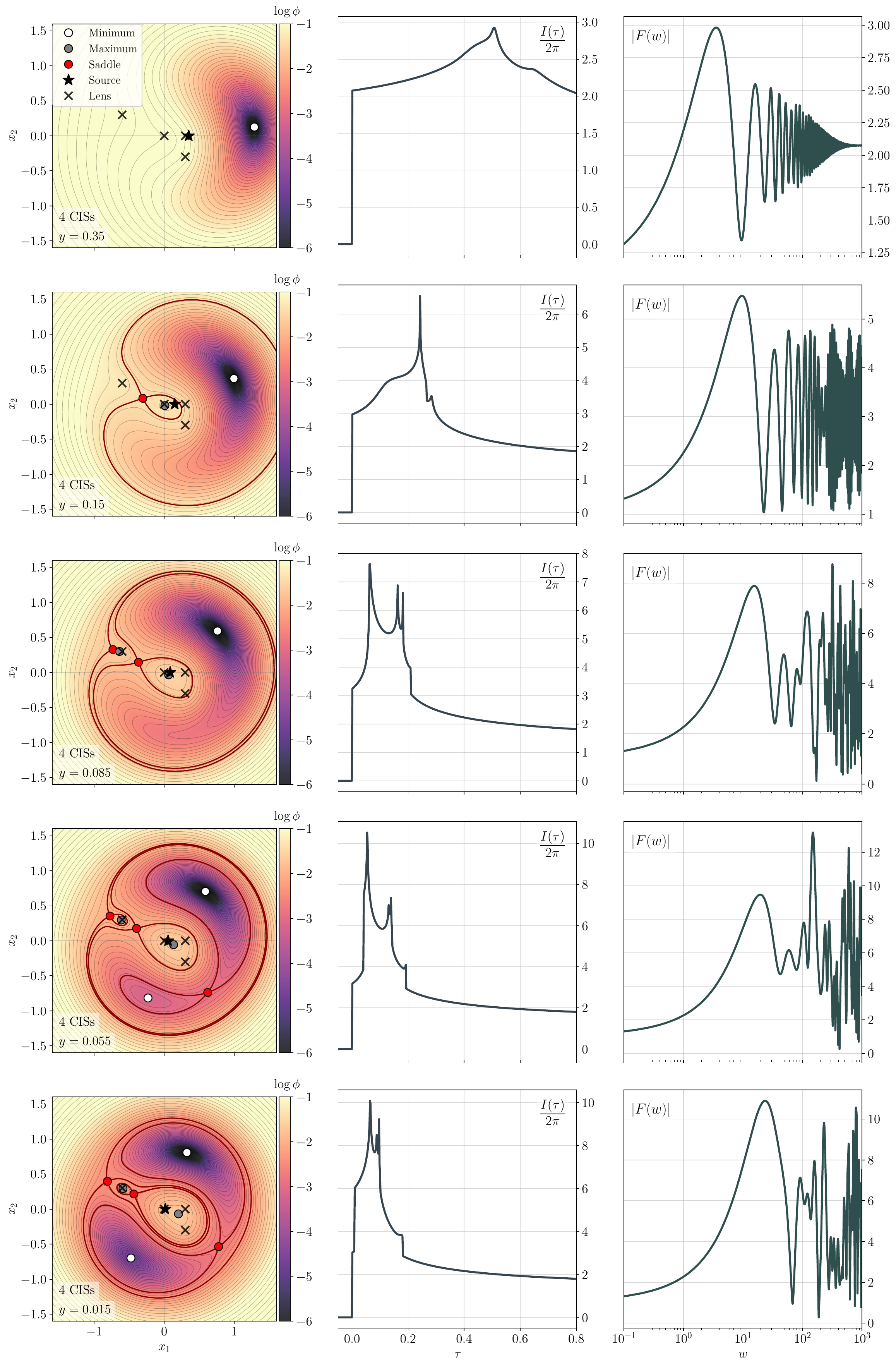}
            \caption{Example of different non-trivial strong lensing configurations. The lens is 
            fixed, a composite lens with four CISs, and the impact parameter is progressively 
            reduced, from top to bottom. The configuration in the lowest panel corresponds to Fig. 
            \ref{fig:simple_contours}. The two lower panels ($y=0.055$ and $y=0.015$) illustrate 
            how two configurations with the same number of critical points may have a different 
            saddle point structure.
            }
            \label{fig:SL_example}
        \end{figure*}

        The steps in the computation for a typical work session can be summarized as follows:
        \begin{enumerate}[I)]
            \item \textit{Choose a lens}. The lens can be chosen from the comprehensive catalog in 
                App. \ref{app:lenses}, or constructed as the combination of any of these lenses. The lenses
                are implemented in Python and C, with a script in place to check that the implementations
                are consistent. In the documentation we provide a guide to implement a new lens model. 
            \item \textit{Compute $I(\tau)$}. After defining the lens and the impact parameter $y$, we can
                choose the appropriate method to compute the time-domain integral. By default, it will be
                precomputed on a grid when we initialize the object and any further calls will evaluate an 
                interpolation function.
            \item \textit{Compute $F(w)$}. Choosing again from the different methods, we can compute the
                amplification factor in the frequency domain from the time domain version. 
                Upon initialization,  the method shown in the example precomputes $F(w)$ in a frequency 
                grid, and the evaluation is performed using an interpolation function. This is crucial 
                for waveform lensing, where the amplification factor must be evaluated in a large number 
                of points. 
            \item \textit{Transform to physical units and lens the waveform}. In general, there is not a 
                one-to-one correspondence between the physical parameters of the lens and its dimensionless
                counterpart used in lensing computations. For instance, the computation shown in the example
                is valid for any values of the lens mass or redshift, but these parameters are only 
                relevant when we want to convert back to physical frequencies. To lens a waveform we must 
                first choose the physical parameters of the lens and the waveform to be lensed. We provide 
                two additional Python modules to facilitate this task, interfacing with 
                \texttt{astropy}\footnote{\url{https://github.com/astropy/astropy}}
                \citep{astropy:2013, astropy:2018, astropy:2022} and 
                \texttt{pycbc}\footnote{\url{https://github.com/gwastro/pycbc}} 
                \cite{Biwer:2018osg, nitz2024pycbc}.
                This allows \glow{} to directly compute lensed waveforms (in time and frequency domain)
                starting from generic lenses. See Fig. \ref{fig:td_waveform} for an example.
        \end{enumerate}    
    
        In the remainder of this Section we will review how the methods discussed in Sec. 
        \ref{sec:time} and \ref{sec:freq} are implemented in \glow{}, focusing on the C version, as well 
        as the code's performance.

    \subsection{Time domain} 
        The time domain module contains five different methods to compute $I(\tau)$:
        \begin{itemize}
            \item \texttt{MultiContour}. This class implements the contour method, outlined in Sec. 
                \ref{sec:contour}, in full generality. It can handle non-axisymmetric lenses with one 
                or multiple images. See Fig. \ref{fig:SL_example} for an application.
            \item \texttt{SingleContour}. This class also implements the contour method, but only in the
                single-image regime. Whenever possible, it uses the angular integration 
                \eqref{eq:sys_ODE_def_nonparam}, that allows for a faster computation in simple scenarios.
                This class should be preferred over \texttt{MultiContour} for weak lensing computations. 
                A complex example is shown in Fig. \ref{fig:WL_example}.
            \item \texttt{SingleIntegral}. Implementation of the method outlined in Sec.    
                \ref{sec:single_integral}. It only supports axisymmetric lenses, with one or multiple 
                images. It is by far the fastest and most reliable method for symmetric lenses. It should
                be always used over the previous methods when dealing with symmetric lenses. 
            \item \texttt{AreaIntegral}. This class implements the method explained in Sec. \ref{sec:grid}.
                It supports arbitrary lenses with multiple images. However, the algorithm is very naive
                and the implementation is not optimized. Its main purpose in the code is to provide an 
                independent cross-check for the other methods. In contrast with all the other methods, it
                does not require the precomputation of the critical points of the Fermat potential, only 
                the global minimum needs to be found. This method should be avoided for anything beyond
                cross-checking. 
            \item \texttt{AnalyticSIS}. Implementation of \eqref{eq:I_sis_analytic}.
        \end{itemize}
        
    \subsection{Frequency domain} 
        The frequency domain module contains four different methods to compute $F(w)$:
        \begin{itemize}
            \item \texttt{FFT}. Computation of the amplification factor using a Fast Fourier Transform (FFT).
                The implementation follows the scheme outlined in Sec. \ref{sec:fft_multigrid}. It is the
                fastest way to compute $F(w)$ for a generic lens. Keep in mind that depending on the input,
                i.e. lens parameters or frequencies requested, some precision parameters may need to be 
                tuned to achieve optimal results. 
            \item \texttt{DirectFT}. Implementation of the Fourier sum presented in \ref{sec:directft}. 
                Since we precompute $I(\tau)$ on a grid and then approximate it by a linear interpolation
                function (by default, but can be changed), this algorithm returns the exact Fourier
                transform. Its main disadvantage with respect to \texttt{FFT} is that each frequency $w$ 
                must be computed individually, making this method usually slower. A clear advantage of 
                this method is that the errors arise only from the insufficient sampling of $I(\tau)$, 
                which can be easily improved. This class can be used for highly precise tasks, without 
                high speed requirements, and as a cross-check of \texttt{FFT}.
            \item \texttt{SemiAnalyticSIS}. Implementation of \eqref{eq:Fw_sis_analytic}.
            \item \texttt{AnalyticPointLens}. Implementation of the analytic formula for the point lens, 
                following the approximations of Sec. \ref{sec:Fw_pl_analytic}.
        \end{itemize}

    \subsection{Performance and precision}
        In this section we will present some tests of the code. All of them have been performed in a laptop 
        with an AMD Ryzen 7 8845HS, using 8 threads, and with the default precision parameters.

        Table \ref{tab:timing_It} contains the initialization times for an SIS lens, using different
        methods in the time domain, both in the strong and the weak lensing regime. Tables 
        \ref{tab:timing_Fw_SL} and \ref{tab:timing_Fw_WL} show the initialization time for the
        amplification factor, for $y=0.3$ and $y=1.2$ respectively. We also include the evaluation
        time for the analytical point lens, using our improved implementation. The tables' three
        columns represent: 
        \begin{enumerate}
            \item Initialization time of $F(w)$, given a time-domain version $I(\tau)$. 
            \item Evaluation time on a frequency grid with $1000$ points.
            \item Total time including the initialization of  $I(\tau)$. The analytical expressions 
                do not require $I(\tau)$ and their initialization time is negligible.
        \end{enumerate}

        For the precision tests, we compared our new numerical methods with the analytical expressions
        available for the SIS and the point lens. We check the precision of $I(\tau)$ for the SIS in Fig. 
        \ref{fig:It_SIS_precision}, and for the amplification factor we test both the SIS in Fig. 
        \ref{fig:Fw_SIS_precision} and the point lens in \ref{fig:Fw_PL_precision}. The default 
        precision parameters are chosen to target a (relative) tolerance of $10^{-4}$ for $I(\tau)$ and 
        $10^{-3}$ for $F(w)$, \emph{after} interpolation. We also show an example with a tolerance
        of $10^{-8}$ with non-default precision parameters. Finally, in Fig. \ref{fig:PL_approx} we 
        compare our, extremely fast, implementation of the analytical amplification factor for the 
        point lens with a naive implementation using arbitrary precision. The switches between the 
        different approximations discussed in Sec. \ref{sec:Fw_pl_analytic} are calibrated to never
        exceed a relative error of $10^{-5}$.

    \begin{table}
        \begin{ruledtabular}
        \begin{tabular}{lll}
            $I(\tau)$ method & $y=0.3$ & $y=1.2$\\\hline
            \texttt{SingleContour} & \rule{1cm}{0.4pt} & 5.5 ms\\
            \texttt{MultiContour} & 16 ms & 13 ms\\
            \texttt{SingleIntegral} & 490 $\mu$s & 480 $\mu$s\\
            \texttt{AnalyticSIS} & 770 $\mu$s & 740 $\mu$s\\
        \end{tabular}
        \end{ruledtabular}
        \caption{Four different computations of the time domain integral for the SIS. The evaluation 
        times correspond to the default initialization time of these objects in the code, which 
        involves computing $I(\tau)$ on a grid with 5000 points. Notice that the 
        \texttt{SingleContour} method can only be used in the single-image regime.}
        \label{tab:timing_It}
    \end{table}

    \begin{table}
        \begin{ruledtabular}
        \begin{tabular}{lllll}
            $F(w)$ method & Initialization & Eval. $w$ & Total\\\hline
            \texttt{FFT} & 1.3 ms & 43 $\mu$s & $<17$ ms\\
            \texttt{DirectFT} & 63 $\mu$s & 4.8 ms & $<21$ ms\\
            \texttt{SemiAnalyticSIS} & \rule{1cm}{0.4pt} & 560 $\mu$s & 560 $\mu$s\\\hline\hline
            \texttt{AnalyticPointLens} & \rule{1cm}{0.4pt} & 77 $\mu$s & 77 $\mu$s
        \end{tabular}
        \end{ruledtabular}
        \caption{Computation of the amplification factor for $y=0.3$, for the SIS (first three 
        rows) and point lens (last row). The total time includes the initialization of $I(\tau)$
        in Table \ref{tab:timing_It}, and we take the slowest method (\texttt{MultiContour}) as an 
        upper limit. For symmetric lenses it would be more appropriate to use 
        \texttt{SingleIntegral}, which would yield a total time of less than 2 ms.}
        \label{tab:timing_Fw_SL}
    \end{table}

    \begin{table}
        \begin{ruledtabular}
        \begin{tabular}{lllll}
            $F(w)$ method & Initialization & Eval. $w$ & Total\\\hline
            \texttt{FFT}& 1.1 ms & 43 $\mu$s & $<14$ ms\\
            \texttt{DirectFT} & 46 $\mu$s & 4.8 ms & $<18$ ms\\
            \texttt{SemiAnalyticSIS} & \rule{1cm}{0.4pt} & 1.3 ms & 1.3 ms\\\hline\hline
            \texttt{AnalyticPointLens} & \rule{1cm}{0.4pt} & 77 $\mu$s & 77 $\mu$s
        \end{tabular}
        \end{ruledtabular}
        \caption{Equivalent of Table \ref{tab:timing_Fw_SL}, for impact parameter
        $y=1.2$.}
        \label{tab:timing_Fw_WL} 
    \end{table}

\section{Conclusions} \label{sec:concl}

Advances in astronomy may soon enable the detection of wave-optics gravitational lensing phenomena, in which the undulatory nature of the signal becomes manifest. Here we have presented numerical methods capable of studying wave-optics lensing phenomena for lenses representing generic matter distributions. After an overview of the formalism, we describe multiple algorithms to compute the time-domain amplification integral, the frequency-domain amplification factor, and efficient analytic methods for simple lenses.
These algorithms have been implemented in a companion software package, \textit{Gravitational lensing of Waves (\glow{})}, which is freely available to the scientific community. \glow{} renders the computation of predictions robust for complex lenses. The code is also fast enough to run on a laptop and perform parameter estimation without interpolating precomputed results. It can also be used in generating template banks to search for lensed GW signals that may be missed by regular search pipelines.

As a tool, \glow{} will enable the investigation of novel lensing phenomena. Wave-optics effects will facilitate testing and characterizing dark-matter structures. GWs detected by current and future facilities will provide stringent constraints on objects above $M\gtrsim 100M_\odot$~\cite{Oguri:2020ldf, GilChoi:2023qrz, Urrutia:2021qak, Fairbairn:2022xln, Tambalo:2022wlm, Zumalacarregui:2024ocb}. Detection of wave-optics lensing signatures seems promising for LISA~\cite{Caliskan:2022hbu, Savastano:2023spl}, especially in the presence of a background lens and for dense dark-matter halos~\cite{Brando:2024inp}. Pulsar-timing arrays hold the promise to detect wave-optics features by galactic-scale lenses, which may enable a detection of the universe's expansion rate~\cite{Jow:2024bwq}.
Similarly, the high-frequency of electromagnetic observations will allow them to probe sub-solar objects, including planets, primordial black holes and compact dark-matter structures~\cite{Ulmer:1994ij,Jow:2020rcy,Tamta:2024pow}. These studies have been largely based on symmetric lenses: \glow{}'s flexibility to incorporate new lens profiles and tackle complex configurations will greatly contribute to this program.

Future developments will expand the capabilities of \glow{}. In the near future, we will expand the catalog of lenses, including non-parametric profiles.
To access increasingly realistic scenarios, we will integrate \glow{} with publicly available lensing codes like \texttt{lenstronomy}\footnote{\url{https://github.com/lenstronomy/lenstronomy}}
\cite{Birrer:2018xgm, Birrer:2021wjl} or 
\texttt{glafic}\footnote{\url{https://github.com/oguri/glafic2}} \cite{Oguri:2010rh}.
To describe wave-optics lensing by the dense stellar fields in lens galaxies (microlensing)~\cite{Diego:2019lcd,Mishra:2021xzz,Shan:2023ngi}, we will further optimize and extend our algorithms to account for large number of lenses.
To describe collective lensing effects, we will develop multi-plane lensing and ray-tracing in the wave-optics regime~\cite{Feldbrugge:2020tti}.
Finally, extensions to \glow{} can address scenarios of new physics, such as tests of general relativity~\cite{Ezquiaga:2020dao,Dalang:2020eaj,Goyal:2023uvm} or time-dependent dark-matter backgrounds~\cite{Jung:2020aem}. 

\begin{acknowledgments}
    We are very grateful to G. Brando, P. Mehta, N. Menadeo, J. Raynaud, S. Singh, D. Yushchenko for their feedback on early versions of the code. We also thank E. Berti, M. Caliskan, M. Cheung, S. Delos, JM Ezquiaga, D. Jow, R. Lo, A. Kumar Mehta, A. Mishra, K. Ng, X. Shan, S. Vegetti, Y. Wang for discussions along the way. 
    Besides the libraries mentioned in the main text, \glow{} also relies on Numpy~\cite{harris2020array}, Scipy~\cite{2020SciPy-NMeth} and 
    Colossus\footnote{\url{https://bitbucket.org/bdiemer/colossus/src/master/}} 
    \cite{Diemer:2017bwl}.
    LC was supported by the Princeton German Department's Summer Work Program.
    HVR is supported by the Spanish Ministry of Universities through a Margarita Salas Fellowship, with funding from the European Union under the NextGenerationEU programme.
\end{acknowledgments}

\clearpage
\appendix
\section{Lens catalog}\label{app:lenses}
In this Appendix, we will provide a detailed catalog of lens models implemented in \glow{}. For each lens, we provide its defining density $\rho(\v{r})$, projected density $\Sigma(\v \xi)$ (obtained using Eq.~\eqref{eq:def_Sigma}) and expression for the lensing potential $\psi(\v x)$ (obtained solving Eq.~\eqref{eq:nabla_psi}). As in the main text, in the following we define $x \equiv | \v{x} |$.

    \subsection{Point lens (PL)}
        The \emph{PL} model is defined by a particle of mass $M$ localized in a point and resulting in an axially symmetric lensing potential. Its density and projected density are
        \begin{subequations}
        \begin{align}
            \rho(\v{r}) &= M\delta^3(\v{r})\ ,\\
            \Sigma(\v{\xi}) &= M \delta^2(\v{\xi})\ . 
        \end{align}
        \end{subequations}
        Defining the Einstein radius as
        \begin{equation}
            R_E \equiv \sqrt{4GM(1+z_L)d_\text{eff}}\ ,
        \end{equation}
        we can write the lensing potential as
        \begin{equation}
            \psi(x) = \psi_0 \log x\,,\qquad
            \psi_0 \equiv \frac{R_E^2}{\xi_0^2}\ .
        \end{equation}
        In this lens model, the so-far unspecified scale $\xi_0$ can be conveniently fixed by setting $\psi_0 = 1$. Note that with this choice the redshifted effective lens mass $M_{Lz}$, defined in Eq.~\eqref{eq:def_MLz}, coincides with the redshifted mass, $M_{Lz} = M (1 + z_L)$.
        
    \subsection{Singular Isothermal Sphere (SIS)}
    The \emph{SIS} is a spherically symmetric lens model, defined by a density decaying as an inverse square power of the radial distance, that is typically used to describe halo profiles. The density and projected density are therefore given by
        \begin{subequations}
        \begin{align}
            \rho(r) &= \frac{\sigma_v^2}{2\pi G\, r^2}\ ,\\
            \Sigma(\xi) &= \frac{\sigma_v^2}{2G\,\xi}\ , 
        \end{align}
        \end{subequations}
    where the parameter $\sigma_v$ is the velocity dispersion of the halo. The lensing potential takes the following simple form
        \begin{equation}
            \psi(x) = \psi_0\,x\,,\qquad
            \psi_0 \equiv \frac{\sigma_v^2}{G \, \Sigma_\text{cr}\,\xi_0}\ .
        \end{equation}
    In this lens model a useful choice for the scale $\xi_0$ is then $\xi_0 = \sqrt{4 G M_{Lz} d_{\rm eff}} \equiv \sigma_v^2 / (G \, \Sigma_{\rm cr})$, such that $\psi_0 = 1$. See e.g.~\cite{Brando:2024inp, Savastano:2023spl} for additional details.
    \subsection{Cored isothermal sphere (CIS)}
    The \emph{CIS} lens is a deformation of the SIS, where the density smoothens at the lens's centre thanks to the presence of a core of radius $r_c$. It is defined as \cite{hinshaw1987gravitational, Flores:1995dc} 
        \begin{subequations}
        \begin{align}
            \rho(r) &= \rho_0\frac{r_c^2}{r^2 + r_c^2}\ ,\\
            \Sigma(\xi) &= \frac{\pi\rho_0r_c^2}{\sqrt{\xi^2+r_c^2}} .
        \end{align}
        \end{subequations}
    Here $\rho_0$ is the central density of the profile. The lensing potential is then obtained as
        \begin{subequations}
        \begin{align}
            \psi(x) 
            &= 
            \psi_0\sqrt{x_c^2 + x^2} \nonumber\\
            &
            \quad 
            +  x_c \psi_0 \log \left(\frac{2x_c}{\sqrt{x_c^2+x^2}+x_c}\right)\ ,\\
            \psi_0 
            &\equiv 
            \frac{2\pi\rho_0r_c^2}{\Sigma_{\rm cr}\xi_0}\ ,\\
            x_c &\equiv \frac{r_c}{\xi_0}\ .
        \end{align}
        \end{subequations}
    The parameter $x_c$ represents a dimensionless core radius and depends on the normalization scale $\xi_0$. 
    The latter quantity can be conveniently fixed by setting $\psi_0 = 1$, as in the previous lens models. This translates into $\xi_0 = 2 \pi \rho_0 r_c^2 / \Sigma_{\rm cr}$. Additional details are given for instance in \cite{Tambalo:2022wlm, Savastano:2023spl}.
    
    \subsection{Truncated singular isothermal sphere (tSIS)}
    The mass enclosed by an SIS profile is logarithmically divergent, given the $\sim 1 / r^{2}$ decay of the density. A truncation of the profile at large radii is then needed. In the \emph{tSIS} lens, the density is truncated at a radius $R$ with a Gaussian 
    cutoff: 
        \begin{subequations}
        \begin{align}
            \rho(r) &= \frac{M_\text{tot}}{2\pi^{3/2}R^3}\left(\frac{r}{R}\right)^{-2}
                    \ee^{-r^2/R^2}\ ,\\
            \Sigma(\xi) &= \frac{M_\text{tot}}{2\pi^{1/2}R^2}\frac{R}{\xi}\,\text{erfc}(\xi/R)\ .
        \end{align}
        \end{subequations}
    Here, $M_{\rm tot}$ is the total mass of the profile and $\text{erfc}(z)$ is the complementary error function. 
    The lensing potential admits an analytic expression as follows:
        \begin{subequations}
        \begin{align}
            \psi(x) 
            &= 
            \psi_0 \, x \, 
            \Bigg\{
                \frac{1}{2u\sqrt{\pi}}
                \left[2\log(u) + E_1(u^2) + \gamma_E\right]\nonumber\\
                &\qquad\qquad
                + \text{erfc}(u) + \frac{1-\text{e}^{-u^2}}{u\sqrt{\pi}}
            \Bigg\}\ ,\\
            u &\equiv x\,\xi_0/R\ ,\\
            \psi_0 &\equiv \frac{M_\text{tot}}{\sqrt{\pi}\Sigma_\text{cr}R\,\xi_0}\ ,
        \end{align}
        \end{subequations}
        where $E_1(z)$ is the exponential integral and $\gamma_E$ is the Euler-Mascheroni constant.
        Again, the most convenient choice for the scale $\xi_0$ is obtained by setting $\psi_0 = 1$.
    
    \subsection{Elliptical Singular Isothermal Sphere (eSIS)}
    The following lens is a phenomenological model for a potential with elliptical symmetry. 
    For simplicity, we added the ellipticity in the lensing potential, rather than the 
    mass density. See \cite{Keeton:2001ss} and references therein for more details on 
    elliptical lens profiles. This lens, that we have (somewhat paradoxically) dubbed 
    elliptical SIS, \emph{eSIS}, is a specific example of the softened power law potential 
    in \cite{Keeton:2001ss}. The lensing potential is 
    \begin{equation}
            \psi(\v x) = \psi_0\,\sqrt{x_1^2 + x_2^2/q^2}\,,\qquad
            \psi_0 \equiv \frac{\sigma_v^2}{G \, \Sigma_\text{cr}\,\xi_0}\ ,
    \end{equation} 
    where $q$ controls the ellipticity of the profile. 
    Different choices for the orientation of the semi-major axis are implemented through 
    a rotation of angle $\alpha$ in the lens plane
    \begin{subequations}
    \begin{align}
        x'_1 &= \cos\alpha\,x_1 - \sin\alpha\,x_2 \,, \\
        x'_2 &= \sin\alpha\,x_1 + \cos\alpha\,x_2 \ .
    \end{align}
    \end{subequations}
    The lensing potential in this more general setting is then given by $\psi_\alpha(\v x) = \psi(x_1', x_2')$.
             
    \subsection{Navarro-Frenk-White (NFW)}
    The \textit{NFW} is the most commonly used spherically symmetric profile to model cold dark-matter halos \cite{Navarro:1996gj} and is defined by the following $\rho(r)$ and $\Sigma(\xi)$
        \begin{subequations}
        \begin{align}
            \rho(r) &= \frac{\rho_s}{(r/r_s)(1+r/r_s)^2}\ ,\\
            \Sigma(\xi) &= 2\rho_sr_s\frac{1-\mathcal{F}(\xi/r_s)}{(\xi/r_s)^2-1}\ ,
        \end{align}
        \end{subequations}
        where
        \begin{equation}
            \mathcal{F}(x) \equiv 
            \begin{cases}
                \displaystyle
                \frac{1}{\sqrt{x^2-1}}\arctan\left(\sqrt{x^2-1}\right)\,, &\ 
                    x>1\\
                \displaystyle
                \frac{1}{\sqrt{1-x^2}}\text{arctanh}\left(\sqrt{1-x^2}\right)\,. &\ 
                    x<1\\
            \end{cases}
        \end{equation}
        The parameter $r_s$ is the so-called scale radius while $\rho_s / 4$ is the density at $r_s$.  
        Solving the projected Poisson's equation one finds the following form for the lensing potential
        \begin{subequations}
        \begin{align}
            \psi(x) 
            &= 
            \frac{1}{2}\psi_0 \big[\log^2(u/2) + (u^2-1)\mathcal{F}^2(u)\big]
            \,,\\
            u &\equiv x\,\xi_0/r_s\ ,\\
            \psi_0 &\equiv \frac{4\rho_s r_s^3}{\Sigma_\text{cr}\xi_0^2}\ .
        \end{align}
        \end{subequations}
        
        As for the lens models discussed previously, a convenient choice is to set $\psi_0 = 1$. However, different normalization choices are often adopted in lensing applications (see e.g.~\cite{Fairbairn:2022xln, Guo:2022dre, Choi:2021bkx}).\vspace{0.5cm} 

\subsection{External Shear and Convergence}
The external effects due to a galaxy or cluster can be represented by constant convergence and shear. This is particularly useful for embedding small-scale lenses like stars or black holes within a galaxy \cite{Mishra:2021xzz}
\begin{equation}
    \psi(x_1,x_2) = \frac{\kappa}{2} (x_1^2 + x_2^2) + \frac{\gamma_1}{2} (x_1^2 - x_2^2) + \gamma_2 x_1 x_2 \ .
\end{equation}
In the above lensing potential, the parameter $\kappa$  is the convergence while $\gamma_1$ and $\gamma_2$ are shear components along $x_1$ and $x_2$ respectively.

\section{Regularization scheme}\label{app:reg}
    The regularizing functions used in \eqref{eq:It_sing} are
    defined as
    \begin{align}
        R_0(\alpha, \beta,\sigma; x)
            &\equiv \frac{\beta\,\Theta(x)}{\left(x^2+(\beta/\alpha)^{2/\sigma}\right)^{\sigma/2}}  \ ,\\
        R_1(\alpha, \beta,\sigma; x)
            &\equiv xR_0(\alpha, \beta, \sigma+1;x)\ ,\\
        R_L(\alpha, \beta; x) 
            &\equiv R_1(\alpha, \beta, 1; x) = \frac{\beta x \,\Theta(x)}{x^2+\beta/\alpha}\ ,\\
        S(A, B; x) 
            &\equiv \frac{AB}{2}\Theta(x)\log\left|\frac{B+x}{B-x}\right|\ ,\\
        S_\text{full}(A, B; x) 
            &\equiv S(A,B;x) - R_L(A, AB^2;x)\ .
    \end{align}
    Their frequency-domain versions are defined as 
    \begin{equation}
        \tilde{R}(w) \equiv -\ii w\int^\infty_{-\infty}\ee^{\ii w\tau}R(\tau)\,\di\tau\ ,
    \end{equation}
    All of them are analytical, and the integrals needed can be found in 
    \cite{gradshteyn2007table},
    \begin{widetext}
    \begin{align}
        \tilde{R}_0(\alpha, \beta, \sigma;\ w) &= 
            \sqrt{\pi}\alpha\Gamma\left(1-\frac{\sigma}{2}\right)
            \left\{\frac{2}{\pi}\left(\frac{wC_0}{2}\right)^{\frac{1+\sigma}{2}}\ee^{-\ii\frac{\pi\sigma}{2}}
            K_\frac{1-\sigma}{2}(wC_0)-\left(\frac{wC_0}{2}\right)\tilde{\mathbb{M}}_\frac{1-\sigma}{2}(wC_0)\right\} \ ,\\
        \tilde{R}_1(\alpha, \beta, \sigma;\ w) &= \frac{2\alpha C_1}{\sqrt{\pi}}\left(\frac{wC_1}{2}\right)^{1+\frac{\sigma}{2}}
            \ee^{-\ii\frac{\pi\sigma}{2}}\Gamma\left(\frac{1-\sigma}{2}\right)K_{1-\frac{\sigma}{2}}(wC_1)\nonumber\\
            &\quad +\frac{\ii\alpha C_1}{1-\sigma}wC_1\left\{1+\sqrt{\pi}\Gamma\left(\frac{3-\sigma}{2}\right)
                \left(\frac{wC_1}{2}\right)\tilde{\mathbb{M}}_{1-\frac{\sigma}{2}}(wC_1)\right\}\ ,\\
        \tilde{R}_L(\alpha, \beta; \ w) &= \frac{\pi}{2}\beta w\,\ee^{-w\sqrt{\beta/\alpha}}
            +\ii \beta \frac{w}{2}\left\{\ee^{-w\sqrt{\beta/\alpha}}\text{Ei}\left(w\sqrt{\frac{\beta}{\alpha}}\right)
                -\ee^{w\sqrt{\beta/\alpha}}E_1\left(w\sqrt{\frac{\beta}{\alpha}}\right)\right\}\ ,\\
        \tilde{S}(A, B;\ w) &= -\ii\frac{\pi}{2}AB\,\ee^{\ii wB}-\ii AB\Big(\cos(wB)\text{si}(wB)
            -\sin(wB)\text{ci}(wB)\Big)\ ,
    \end{align}
    \end{widetext}
    with $C_0\equiv (\beta/\alpha)^{1/\sigma}$, $C_1\equiv (\beta/\alpha)^{1/(\sigma+1)}$
    and $\tilde{\mathbb{M}}_\nu(z)\equiv \left(2/z\right)^\nu\mathbb{M}_\nu(z)$.
    The following special functions have been used:
    \begin{align*}
        K_\nu(z)\quad : \quad 
            &\text{Irregular modified Bessel function}\\
            &\text{\cite[\href{http://dlmf.nist.gov/10.25}{(10.25)}]{NIST:DLMF}}\\
        \mathbb{M}_\nu(z)\quad : \quad 
            &\text{Modified Struve function}\\
            &\text{\cite[\href{http://dlmf.nist.gov/11.2.E6}{(11.2.6)}]{NIST:DLMF}}\\
        E_1(z),\ \text{Ei}(z)\quad : \quad 
            &\text{Exponential integrals}\\
            &\text{\cite[\href{http://dlmf.nist.gov/6.2.E1}{(6.2.1)}]{NIST:DLMF}
            and \cite[\href{http://dlmf.nist.gov/6.2.E5}{(6.2.5)}]{NIST:DLMF}}\\
        \text{si}(z),\ \text{ci}(z)\quad : \quad 
            &\text{Sine and cosine integrals}\\
            &\text{\cite[\href{http://dlmf.nist.gov/6.2.E10}{(6.2.10)}]{NIST:DLMF}
            and \cite[\href{http://dlmf.nist.gov/6.2.E11}{(6.2.11)}]{NIST:DLMF}}
    \end{align*}

\clearpage
\newpage

\onecolumngrid
\section{Code precision}\label{sec:code_prec}
    \begin{figure*}[h!]
        \includegraphics[scale=0.7]{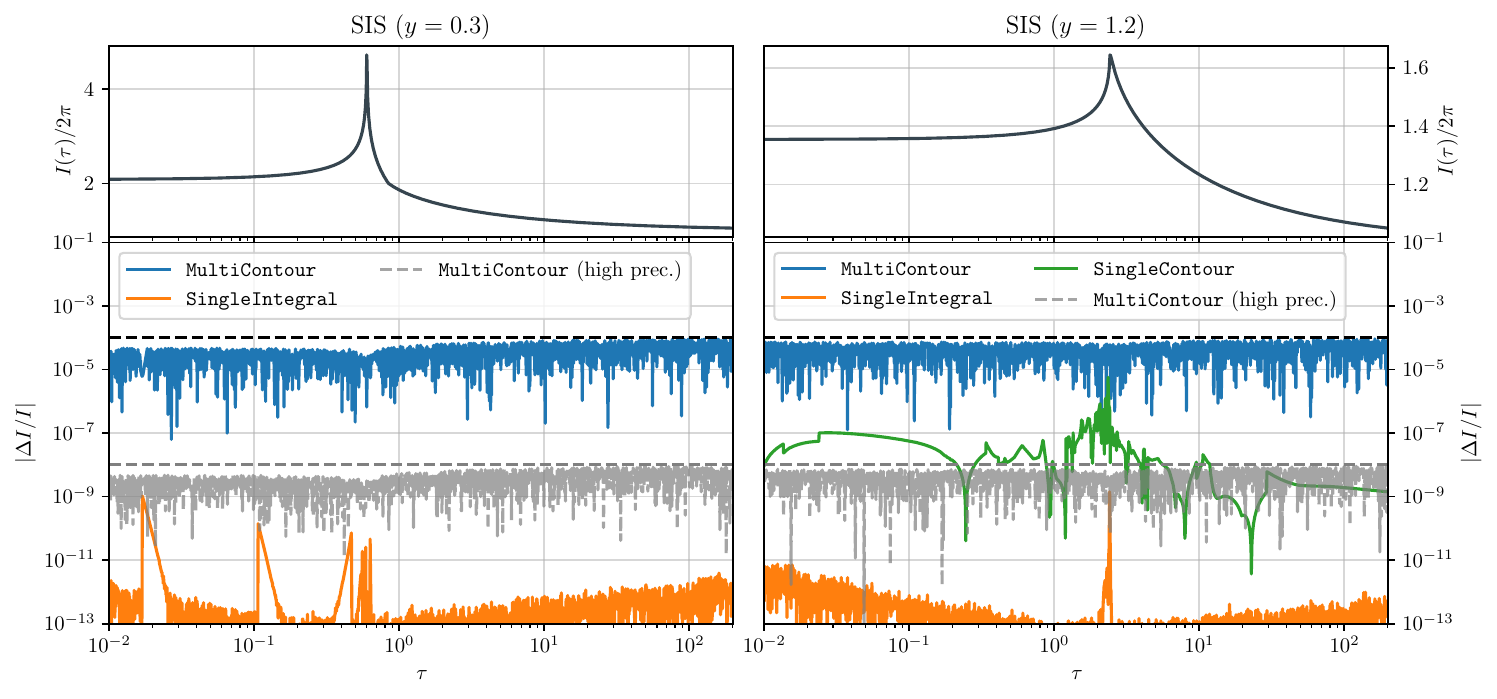}
        \caption{Time domain integral for the SIS, in the strong and weak lensing regimes. In the
        lower panel we show the relative difference with respect to the analytical expression
        \eqref{eq:I_sis_analytic}. In this case, all the points have been computed exactly, i.e.
        without interpolation. The curve labelled as (high prec.) shows how an appropriate tuning
        of the precision parameters can easily increase the precision, without a significant impact
        on the performance. All the other curves use the default precision parameters, chosen
        to ensure a relative tolerance of $10^{-4}$. Note that, due to the simplicity of the SIS
        potential, the \texttt{SingleIntegral} method significantly overperforms in this case.}
        \label{fig:It_SIS_precision}
        \vspace*{5mm}
        
        \includegraphics[scale=0.7]{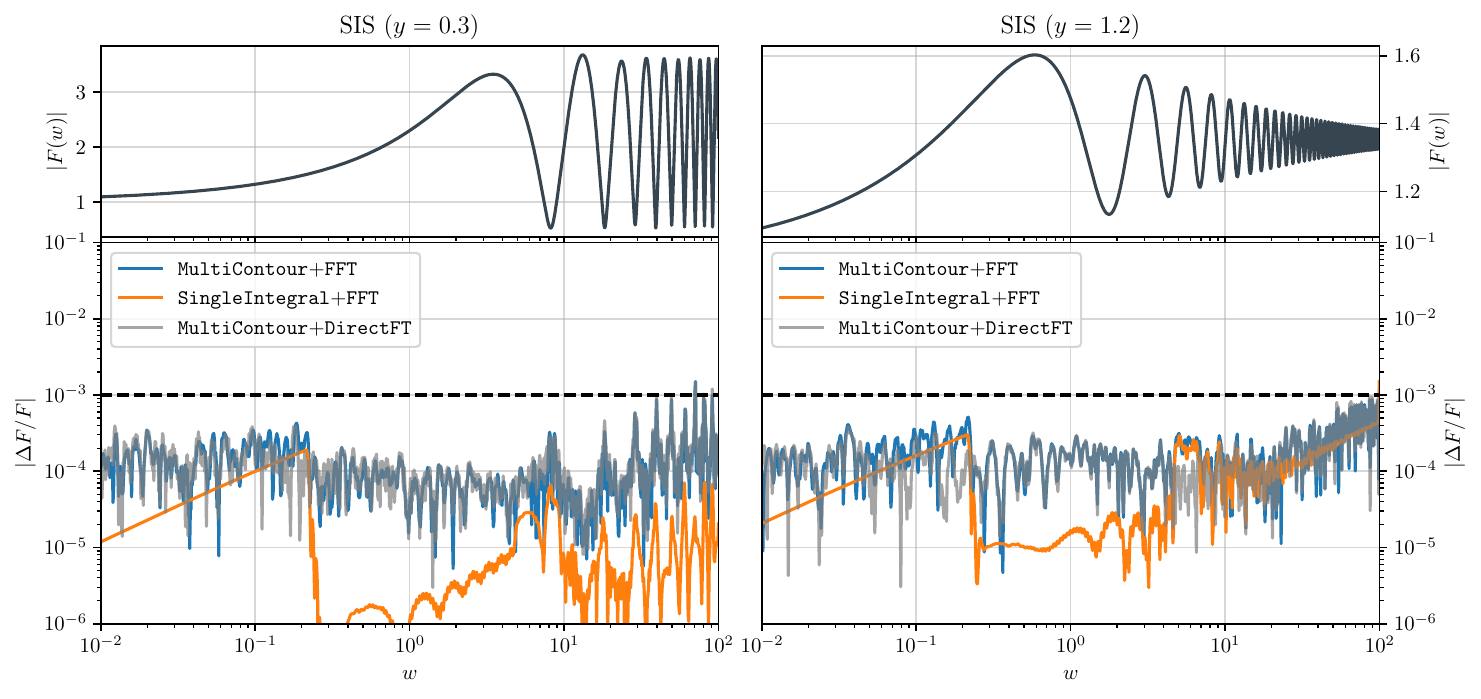}
        \caption{Amplification factor for the SIS, in the strong and weak lensing regimes. In the
        lower panel we show the relative difference with respect to the semianalytic expression
        in \eqref{eq:Fw_sis_analytic}, using the default parameters in the code. It is important 
        to note that these are the results \emph{after} interpolating $F(w)$ on a grid, using cubic
        interpolation. The \texttt{MultiContour} results are dominated by the inaccuracies in the
        computation of the time domain integral, see Fig. \ref{fig:It_SIS_precision}, while the
        errors for \texttt{SingleIntegral} are dominated by the inaccuracies introduced by the
        FFT.}
        \label{fig:Fw_SIS_precision}
    \end{figure*}

    \begin{figure*}[h!]
        \includegraphics[scale=0.7]{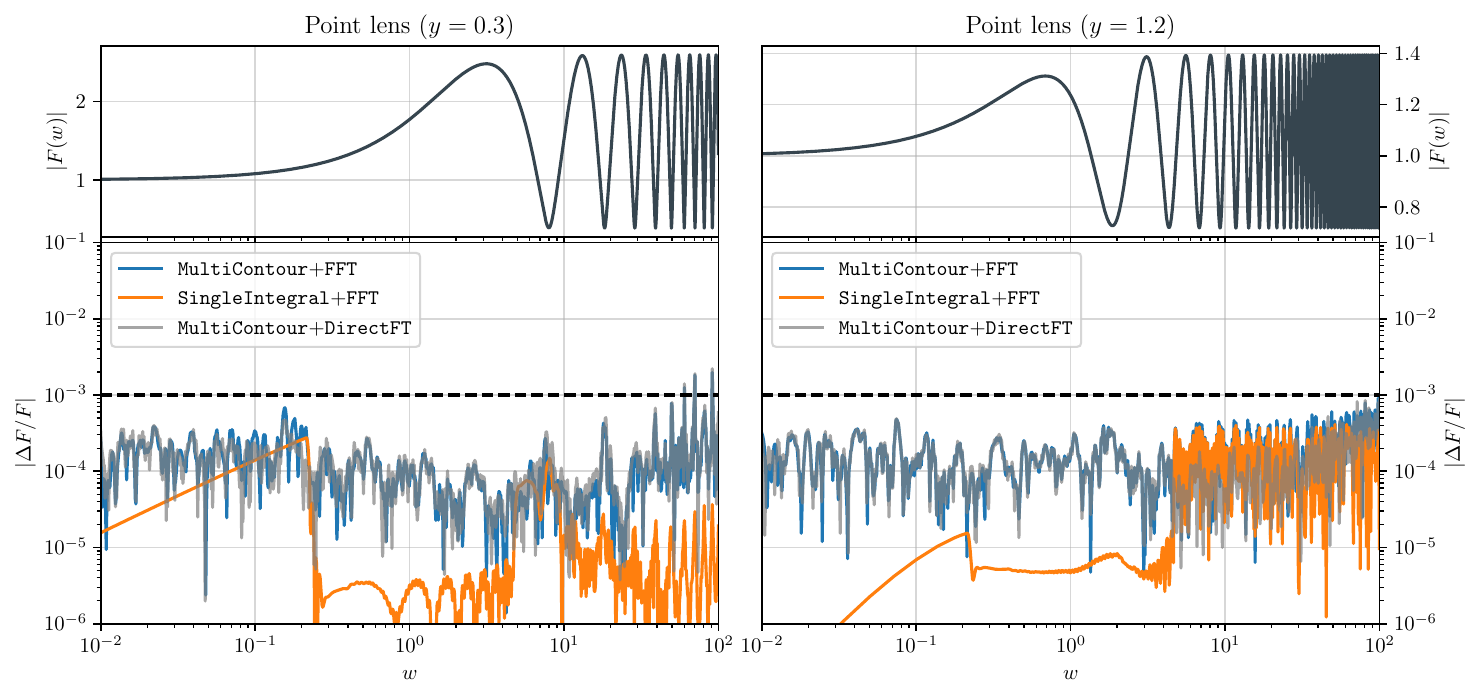}
        \caption{Amplification factor for the point lens, with two different impact parameters. In the
        lower panel we show the relative difference with respect to the analytic expression
        in \eqref{eq:Fw_pl_analytic}, using the default parameters in the code. It is important 
        to note that this are the results \emph{after} interpolating $F(w)$ on a grid, using cubic
        interpolation.}
        \label{fig:Fw_PL_precision}
        \vspace*{15mm}
    
        \includegraphics[scale=0.75]{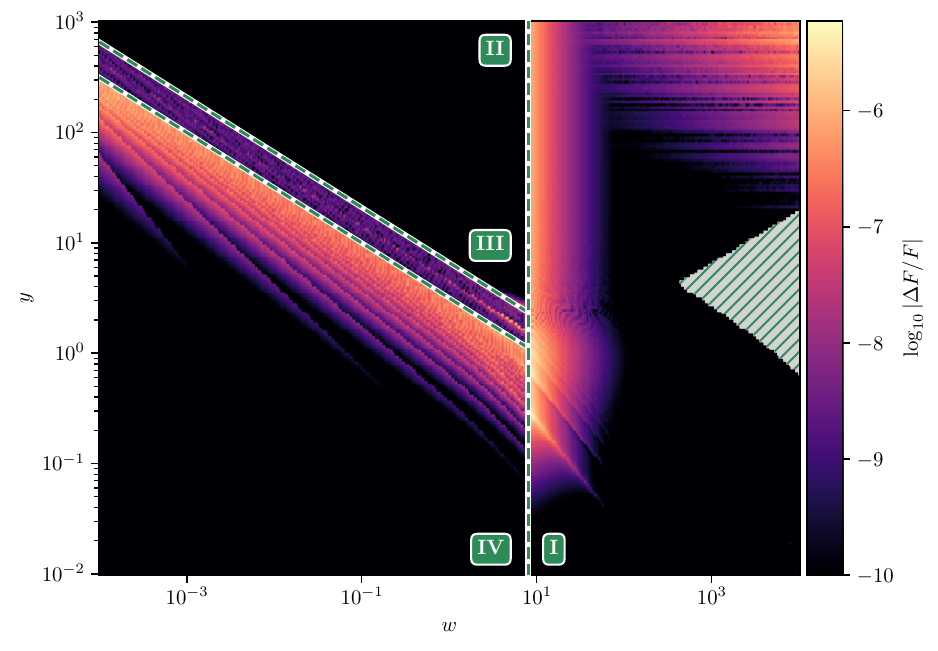}
        \caption{Relative errors on the analytical point lens implementation. We compare the
        implementation based on the approximations discussed in Sec. \ref{sec:Fw_pl_analytic} (and
        highlighted in the figure), with
        a classical implementation using arbitrary precision based on \texttt{mpmath} and also 
        available in our code. The switches between the different approximations are calibrated
        to never exceed a tolerance of $10^{-5}$. The maximum difference in this particular grid
        of $400\times 200$ points is $5\times 10^{-6}$. In the hatched region, the computation 
        using arbitrary precision becomes prohibitively expensive. Finally, the errors in the upper
        right corner arise from numerical errors not in $\tensor[_1]{F}{_1}$, but in the other 
        pieces of the amplification factor. Although $F(w)$ is essentially 1 in this corner, the
        geometric optics approximation excels in this region and it can be used to increase the
        precision further, if needed.}
        \label{fig:PL_approx}
    \end{figure*}
    
\clearpage
\twocolumngrid

\newpage
\bibliographystyle{apsrev4-1}
\bibliography{gw_lensing}

\end{document}